\UseRawInputEncoding
\documentclass[
aps,pra,tightenlines,10pt,notitlepage,nofootinbib,twocolumn,superscriptaddress
]{revtex4-2}
\usepackage[english]{babel}
\usepackage{svg}
\usepackage{graphicx} % Required for inserting images
\usepackage{amsfonts}
\usepackage{amsmath}
\usepackage{amsthm}
\usepackage{amssymb}
\usepackage[bb=boondox]{mathalfa}
\usepackage{nicematrix}
\usepackage{amsmath,epsfig,amssymb}
\usepackage{bbm}
\usepackage{bm}
\usepackage{braket}
\usepackage{mathtools}
\usepackage[T1]{fontenc}

\newcommand{\Ftwo}[0]{\mathbb{F}_2}

\newcommand{\txp}[0]{\widetilde{(x^p)}}
\newcommand{\typ}[0]{\widetilde{(y^p)}}

\newtheorem{lemma}{Lemma}
\newtheorem{theorem}{Theorem}
\newtheorem{definition}{Definition}

\newtheorem{corollary}{Corollary}

\usepackage{float}
\makeatletter
\let\newfloat\newfloat@ltx
\makeatother
\usepackage{algorithm}
\usepackage[noend]{algpseudocode}
\makeatletter
\def\BState{\State\hskip-\ALG@thistlm}
\makeatother

\renewcommand{\S}{\mathcal{S}}

\usepackage[colorlinks=true, citecolor=blue, linkcolor=red!80!black,urlcolor=green!80!black,breaklinks=true]{hyperref}

\graphicspath{ {figures/} }

\begin{document}

\title{Satisfying Quantum Codes:\\Physics-Informed and Hardware-Aware Code Design with SAT Solvers}

\author{Ben DalFavero}
\affiliation{Department of Computational Mathematics, Science, and Engineering, Michigan State University, East Lansing, MI 48823, USA}
\affiliation{Johns Hopkins Applied Physics Laboratory, Laurel, MD 20723, USA}

\author{William M.  Watkins}
\affiliation{William H. Miller III Department of Physics \& Astronomy, Johns Hopkins University, Baltimore, MD 21218, USA}

\author{Margarite L. LaBorde}
\affiliation{Johns Hopkins Applied Physics Laboratory, Laurel, MD 20723, USA}

\author{Vincent Russo}
\affiliation{Unitary Foundation, 315 Montogomery St, Fl 10 San Francisco, California 94104, USA}

\author{Ethan Egger}
\affiliation{Center for Quantum Information and Control, University of New Mexico, Albuquerque, NM 87131, USA}

\author{Gregory Quiroz}
\affiliation{Johns Hopkins Applied Physics Laboratory, Laurel, MD 20723, USA}
\affiliation{William H. Miller III Department of Physics \& Astronomy, Johns Hopkins University, Baltimore, MD 21218, USA}

\author{Ryan LaRose}
\thanks{Corresponding author:  \href{rmlarose@msu.edu}{rmlarose@msu.edu}}
\affiliation{Department of Computational Mathematics, Science, and Engineering, Michigan State University, East Lansing, MI 48823, USA}
\affiliation{Department of Electrical and Computer Engineering, Michigan State University, East Lansing, MI 48823, USA}
\affiliation{Department of Physics and Astronomy, Michigan State University, East Lansing, MI 48823, USA}
\affiliation{Center for Quantum Computing, Science, and Engineering, Michigan State University, East Lansing, MI 48823, USA}

\begin{abstract}
    Although quantum error correction is widely believed to be necessary for impactful applications of quantum computers, the design of quantum error correction codes is largely done by hand, without respect to problem or hardware constraints. In this work, we present a highly general and flexible framework for the computational design of both physics-inspired and hardware-aware quantum codes. To do so, we formulate code design as a Boolean satisfiability (SAT) problem and show how to incorporate all required error correction criteria. We prove that code design is NP-complete, ruling out any efficient algorithm for designing codes in general. Nonetheless, we show that state-of-the-art SAT solvers are able to effectively find solutions for many practical problems. Notably, we are able to design physics-inspired codes with up to 100 physical qubits in minutes, and we design new hardware-aware codes for biased noise which have a lower logical error rate than state-of-the-art surface codes.  
\end{abstract}

\maketitle

% \tableofcontents

\section{Introduction} \label{sec:intro}

It is widely accepted that quantum error correction will be necessary to implement impactful applications on quantum computers. 
This need has motivated substantial effort toward designing quantum codes with lower resource overheads and greater resilience to realistic noise. 
Existing analytical techniques often search for quantum codes from their classical counterparts~\cite{calderbank_good_1996, cross_codeword_2009, bravyi_high-threshold_2024} or using inspiration from mathematics and physics \cite{kitaev_classical_2002,Hastings2021dynamically}. The impact of this search is reflected in the steady reduction of resource requirements for error-corrected implementations of Shor's algorithm~\cite{gidney2025factor2048bitrsa,Fowler2012Surface,Cody2012layered,ogorman2017magic,gheorghiu2019benchmarking,Gidney2021howtofactorbit,dallairedemers2026brace,zhou2025resource,cain2026shorsalgorithmpossible10000}, as well as in key experimental milestones, including demonstrations of break-even and sub-threshold logical memories~\cite{google2023suppressing, ni2023beating, acharya_quantum_2025, he2025qec, brock2025quantum} and logical gate operations~\cite{evered_high-fidelity_2023, bluvstein_logical_2024,Self2024, He2025, paetznick_improved_2026, jin2026icebergtipcocompilationquantum,vezvaee_demonstration_2026}. However, although new codes are continuously being developed, the design of good quantum error correcting codes has proven to be difficult.  There has not been a clear improvement over the surface code in the more than three decades since it was introduced. Designing good quantum codes appears to be hard in general~\cite{kapshikar2023}, even more so when trying to incorporate real-time error rates, qubit inhomogeneity, and other practical considerations about quantum hardware. 

In this work, we develop a computational framework to design quantum codes that is capable of overcoming these challenges. Specifically, we formulate quantum code design as a satisfiability (SAT) problem and leverage state-of-the-art SAT solvers to design novel quantum codes. Our SAT formulation is extremely general and is able to incorporate problem symmetries, which we refer to as physics-inspired code design, as well as detailed noise models, which we refer to as hardware-aware code design. Through this framework, we design quantum codes which demonstrate superior accuracy when computing expectation values in quantum error detection experiments, and quantum codes that have lower logical error rate than the XZZX surface code for biased noise models. We demonstrate the practical scalability of this approach by designing novel codes with over 100 physical qubits in minutes to hours, depending on the application, on a laptop. 

In addition to these findings, we study the complexity of designing codes in general and prove that this is an NP-complete problem by a reduction from the set cover problem. Beyond this worst-case complexity, we analyze the average-case runtime for relevant problem instances and characterize a phase transition in the probability of finding a satisfiable solution, as well as the runtime, for practical codes. In short, we present both a general theoretical framework for quantum code design with rigorous complexity results and average-case performance, in addition to practical tools for quantum code design used to find novel codes with superior performance.

While there is prior work in the computational design of quantum codes, our work significantly extends current state-of-the-art methods in several ways. Notably, our framework is, as mentioned, extremely general and capable of designing new codes from existing problem symmetries and/or hardware-tailored noise models, in addition to designing new codes without any additional constraints. We also exceed the scale of state-of-the-art methods. These methods include Ref.~\cite{olle_simultaneous_2024}, in which reinforcement learning was used to discover codes with constrained connectivity on up to 25 physical qubits, and Ref.~\cite{guerrero_game-theoretic_2026}, in which a multi-agent system was used to optimize quantum codes and enable discovery of a $[\![100,50,4]\!]$ code in under an hour. Using our SAT formulation, we can design  a $[\![100,50,4]\!]$ code in under five minutes of CPU time, in addition to designing novel physics-inspired codes which start from the symmetries of a 100 qubit Fermi-Hubbard model in minutes to hours, depending on the application. Recently, SAT has been used in Ref.~\cite{koh_entangling_2026} to discover so-called phantom codes. Our work enables code discovery as a special case, but we also enable code extension, physics-inspired code design, and hardware-aware code design in a highly general, flexible, and performant framework. Also, SAT has been used in Ref.~\cite{peham_automated_2025} to find state preparation circuits for CSS codes, and in Ref.~\cite{ehatamm2026endtoendformalizationquantumerror} to certify distances in generated quantum codes. Our work again extends this by incorporating physical symmetries, hardware information, and error correction criteria to enable end-to-end quantum code design, and we demonstrate the advantages of this through benchmarks in higher fidelity error-detected quantum algorithms and lower logical error rates in error-corrected memory experiments. 

The structure of the paper is as follows. In Sec.~\ref{sec:main_results}, we summarize  the main results of our work. In Sec.~\ref{sec:preliminaries}, we provide background on quantum error correction codes, satisfiability, and our notation. In
Sec.~\ref{sec:code-design}, we define the problem of quantum code design and discuss particular cases which we focus on --- code discovery and code extension. In Sec.~\ref{sec:complexity}, we prove that the problem of code design is NP-complete using a reduction from the set cover problem. In Sec.~\ref{sec:sat}, we formulate the quantum code design problem in terms of Boolean satisfiability (SAT), showing how to represent error correction criteria in SAT clauses. In Sec.~\ref{sec:phase_transition}, we analyze the SAT--UNSAT transition in the average-case complexity of random instances of the code design problem. In
Sec.~\ref{sec:physics-informed-code-design}, we demonstrate our formalism by designing physics-inspired codes, in particular novel codes stemming from the Fermi-Hubbard model with up to 100 physical qubits. In Sec.~\ref{sec:hardware-aware-code-design}, we demonstrate our formalism by designing hardware-aware codes, notably designing novel surface codes which outperform the XZZX surface code for biased noise.

\section{Main Results\label{sec:main_results}}

\begin{figure*}
    \centering
    \includegraphics[width=\linewidth]{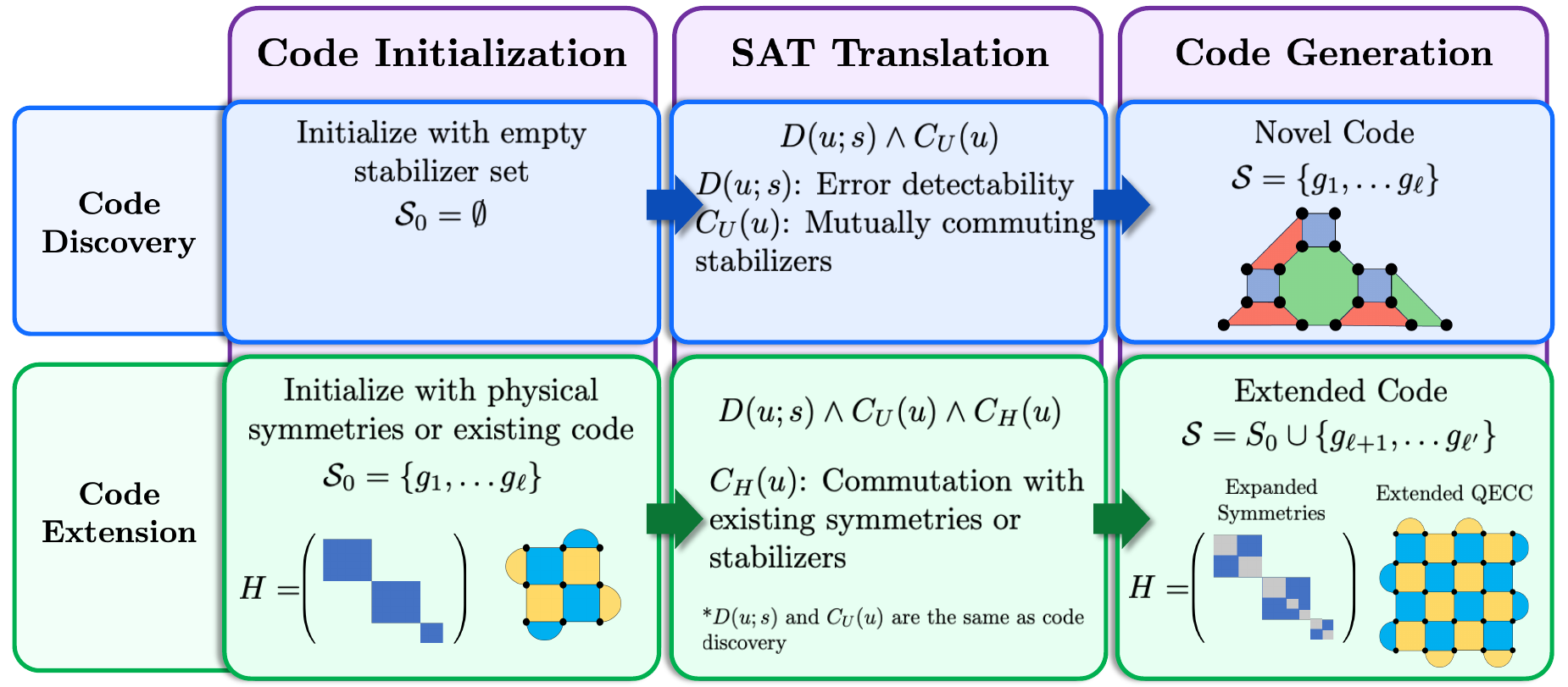}
    \caption{Overview of our quantum code design framework. Our framework enables two categories of code design: code discovery and code extension. In the former, the code is initialized with an empty stabilizer set, while in the latter the initial set $\mathcal{S}_0$ is determined by a predetermined code or a set of physics-informed symmetries captured within the Hamiltonian $H$. Our SAT-based framework enables translation between both code design scenarios through the symplectic form, which encodes information about both the error set $\mathcal{E}$ and stabilizers/symmetries. Solutions to the SAT problem lead to novel codes ``from scratch'' or extended codes that contain the initial stabilizers. Both code discovery and code design incorporate a set of errors $\mathcal{E}$ to be detectable/correctable, enabling hardware-aware code design.}
    \label{fig:overview}
\end{figure*}

In this section we provide a summary of our main results of this work. An overview of our code design framework is shown in Fig.~\ref{fig:overview}. We formally define the problem of quantum code design in Def.~\ref{def:code_design-non-decision} and provide existence proofs for important cases in Sec.~\ref{sec:code-design}. These cases are \textit{code discovery}, in which we start from an empty initial stabilizer group, and \textit{code extension}, in which we start from a non-empty initial stabilizer group. Code discovery is concerned with the generation of new quantum codes, while code extension is concerned with building a code from an existing set of symmetries, either from the physics of a particular problem or from an initial code. We refer to the former case of starting from a set of problem symmetries as \textit{physics-inspired code design}. Both code discovery and code extension can incorporate physical noise models of devices, and we refer to this case as \textit{hardware-aware code design}. Building on the notion of classical code extension, we define extensions of quantum codes in Def.~\ref{def:quantum-code-extension} and prove that distance-increasing extensions exist for quantum codes,
stated in Theorem~\ref{thm:existence-of-distance-increasing-extensions}. We emphasize that a distance-increasing extension $C'$ of a code $C$ satisfies  $C \subseteq C'$ in the sense that $C'$ adds additional qubits and stabilizers to the initial set of stabilizers to increase the distance. We discuss particular examples including extensions with mutations (i.e., in which some initial stabilizers are changed) and extensions without mutations (i.e., in which stabilizers are unchanged). Interestingly, both examples exist for quantum low density parity check (QLDPC) codes. Our work provides a general framework to produce code extensions beyond these few examples.

Next, we analyze the general complexity of the quantum code design problem, and prove the following:
\begin{theorem}[Informal]
    The problem of designing a quantum code is NP-complete. 
\end{theorem}
\noindent The formal version and proof of this statement is given in Theorem~\ref{thm:np-complete}. Our proof uses a reduction from $k$-\textsc{set cover}, one of Karp's original NP-complete problems~\cite{karp_reducibility_1972}. This rules out an efficient solution for designing quantum codes in the general case, and provides a theoretical foundation to prior work in computational code design. 
Nonetheless, in the same way that decoding is NP-hard (in fact, \#P-complete)~\cite{Iyer_Poulin_2015} yet many algorithms have been found for practical cases, we next explore algorithms for solving the quantum code design problem in practically-relevant cases. To do so, we formulate the quantum code design problem as a SAT instance in Sec.~\ref{sec:sat}. This works by writing conditions on stabilizer codes including commutativity and error correction/detection of desired errors as Boolean functions, then creating an overall Boolean formula in conjunctive normal form. This formulation is very general and encompasses code discovery, physics-informed code design, and hardware-aware code design. 

The SAT formulation allows us to apply state-of-the-art SAT solvers~\cite{ignatiev_pysat_2018, ignatiev_towards_2024} for code design and also to analyze the satisfiability probability and the practical runtime of this approach. We do this in Sec.~\ref{sec:phase_transition} in which we empirically find a phase transition in the satisfiability probability, and in average-case time complexity, for instances with randomly-drawn sets of errors. This phase transition is illustrated in Fig.~\ref{fig:phase_transition} where the order parameter is the number of errors per stabilizer. Additionally, we compute a phase diagram for distance-three codes, shown in Fig.~\ref{fig:phase_transition_diagram}. Here, while the Gilbert-Varshamov and Singleton bounds constrain the low- and high-rate regions, respectively, the fitted SAT--UNSAT transition provides an empirical prediction of code existence in the contested region. Generally, the SAT formulation provides a rigorous and powerful framework to prove the (un)satisfiability of a given set of code parameters.

We then apply our code design formalism to creating novel physics-inspired codes in Sec.~\ref{sec:physics-informed-code-design}. Here, we demonstrate our framework by starting from the symmetries of the Fermi-Hubbard Hamiltonian and augmenting them to correct certain errors. We demonstrate that this improves the accuracy of expectation values computed in noisy simulation in Fig.~\ref{fig:fh-error-detection}. Next, we show that we scale this approach and find physics-informed code extensions for an $n = 100$ Fermi-Hubbard Hamiltonian in minutes to hours, shown in Fig.~\ref{fig:fh-extension-time-n48-100}. This significantly extends the problem-size of prior work in state-of-the-art computational code design mentioned in Sec.~\ref{sec:intro}.

Last, we apply our code design formalism to creating novel hardware-aware codes in Sec.~\ref{sec:hardware-aware-code-design}. We first demonstrate this by augmenting the physics-informed code design problem for the Fermi-Hubbard model with biased noise. In this case, our SAT formulation finds novel codes that are tailored to the hardware noise model, and we show these lead to reduced undetected error rates in noisy simulation, shown in Fig.~\ref{fig:bias-tailoring}. Next, we demonstrate our framework by discovering novel surface codes for biased noise models which have a lower logical error rate than current state of the art, under a maximum likelihood decoder. Specifically, we design novel surface codes using our formalism which have logical error rates lower than the XZZX surface code, which is tailored for biased noise. These results, shown in Fig.~\ref{fig:ler-vs-eta-heatmap}, further demonstrate the utility of our code design framework as a tool capable of accelerating the timeline to an experimental demonstration of a large-scale error corrected quantum computer.

\section{Preliminaries\label{sec:preliminaries}}

We begin by reviewing  foundational concepts that are relevant to the development and discussion presented in this work. In particular, we provide an overview of stabilizer codes and describe their representation in binary symplectic form, which will be used throughout our analysis. We also introduce the notation and basic definitions associated with SAT problems that are needed to formulate the code-design problems considered in subsequent sections.

\subsection{Stabilizer Codes}
A quantum stabilizer code is defined by a stabilizer group
\begin{equation} \label{eqn:stabilizer-code-generators}
    \S = \langle g_1, \ldots, g_{n-k} \rangle
\end{equation}
where operators $g_i$ are the stabilizer generators of the code. Throughout the paper we assume all codes are stabilizer codes unless otherwise noted.
The triple $[\![n,k,d]\!]$ refers to a quantum code that encodes $k$ logical qubits into $n$ physical qubits and has distance $d$. When the distance of a code is not known or is unimportant, the double $[\![n,k]\!]$ is used to specify the code. The same three parameters for a classical code are presented as $[n,k,d]$. It is desirable for codes to have both a high encoding rate $k / n$ and high distance $d$. Intuitively, the encoding rate constitutes the number of logical qubits encoded per physical qubit, and the distance defines how many errors it takes to map one codeword onto another. A distance $d$ code can correct arbitrary single-qubit errors on up to $t = \lfloor (d - 1) / 2 \rfloor$ physical qubits. 

The binary symplectic representation maps the $n$-qubit Pauli group $\mathcal{P}_n$ to the binary symplectic space $\Ftwo^{2n}$, \textit{i.e.}, the $2n$-dimensional vector space over the field of integers modulo 2.
In this representation, a Pauli string $P \in \mathcal{P}_n$ is represented as the binary vector $P \cong u \equiv [\vec{x} | \vec{z}]$ with $\vec{x}, \vec{z} \in \Ftwo^n$, where 
\begin{equation}
    \begin{aligned}
    x_i &= 0, & z_i &= 0, && \text{if } P_i = I, \\
    x_i &= 1, & z_i &= 0, && \text{if } P_i = X, \\
    x_i &= 1, & z_i &= 1, && \text{if } P_i = Y, \\
    x_i &= 0, & z_i &= 1, && \text{if } P_i = Z.
    \end{aligned}
\end{equation}
For $P_1, P_2 \in \mathcal{P}_n$, let $P_1 \cong u_1$ and $P_2 \cong u_2$. The product of two Pauli strings is the bitwise exclusive-OR (XOR) of their binary symplectic representations:
\begin{equation}
    P_1 P_2 \cong u_1 \oplus u_2 .
\end{equation}
Whether $P_1$ and $P_2$ commute can be tested using a bilinear form. Let
\begin{equation}
    \Lambda =
    \begin{bmatrix}
        \mathbb{0}_{n \times n} & \mathbb{I}_{n \times n} \\
        \mathbb{I}_{n \times n} & \mathbb{0}_{n \times n}\\
    \end{bmatrix}
\end{equation}
where $\mathbb{I}_{n \times n}$ is the $n \times n$ identity matrix and $\mathbb{0}_{n \times n}$ is the $n \times n$ matrix of all zeros. Using this matrix, we define the symplectic inner product
\begin{equation}
    \langle u_1, u_2 \rangle \coloneq u_1^T \Lambda u_2.
\end{equation}
The symplectic inner product serves as a commutation test since
\begin{equation}
    \langle u_1, u_2 \rangle = \begin{cases}
        0 \text{    if $[P_1, P_2] = 0$}, \\
        1 \text{    if $\{P_1, P_2\} = 0$.}
    \end{cases}
\end{equation}
% $P_1$ and $P_2$ commute if $u_1^T \Lambda u_2 = 0$, and they anticommute if $u_1^T \Lambda u_2 = 1$.
%
The check matrix of a stabilizer code is the matrix that has rows made up of the binary symplectic representations of the code's stabilizer generators, \textit{i.e.}, the check matrix $H$ of the stabilizer code~given in~\eqref{eqn:stabilizer-code-generators} is
\begin{equation}
H =
\begin{bmatrix}
- & u_1 & - \\
  & \vdots &   \\
- & u_{n-k} & -
\end{bmatrix},
\end{equation}
where $g_i \cong u_i$ and the code has stabilizer generators $g_1, \ldots, g_{n-k}$. 

\subsection{Boolean Satisfiability}
SAT problems are defined in terms of $N$ Boolean variables and a set of $M$ constraints where each constraint takes the special form of a clause. In the SAT formulation of code design (Sec.~\ref{sec:sat}), we use the symbols $\wedge$, $\vee$, and $\oplus$ to denote logical conjunction, logical disjunction, and exclusive-OR, respectively. SAT instances are commonly expressed in conjunctive normal form (CNF), in which a formula over Boolean variables $x_1, \ldots, x_N$ is written as a conjunction of clauses. Each clause is a disjunction of literals, where a literal is either a variable or its negation. For example, the formula
\begin{equation}
    (x_1 \vee x_5 \vee \neg x_9) \wedge (\neg x_2 \vee x_{12}) \wedge (x_2)
\end{equation}
is in CNF. Moreover, the conjunction of two or more CNF formulas also yields a CNF formula. 
%Throughout this paper, \texttt{monospace font} is used for the names of Boolean functions.

\section{Quantum Code Design} \label{sec:code-design}

Designing a quantum code entails building a stabilizer group which satisfies a certain set of criteria. Nominally, the main criterion is the set of errors to correct, but additional criteria can include restrictions on minimum or maximum stabilizer weight, (geometric) locality, code distance, and similar properties. Formally, we define the quantum code design problem as follows.

\begin{definition}[$k$-\textsc{code design}]
    \label{def:code_design-non-decision}
    Given a set of $n$-qubit Pauli operators $\Sigma \subseteq \mathcal{P}_n$, 
    a set of $n$-qubit Pauli errors $\mathcal{E}$, 
    and an integer $k$, 
    find a subset of at most $n-k$ linearly independent operators from $\Sigma$ 
    that generates the stabilizer group of an $[\![n,k]\!]$ stabilizer code detecting all errors in $\mathcal{E}$, that is, every $E\in\mathcal{E}$ either has a non-zero syndrome or belongs to the stabilizer group.
\end{definition}

We distinguish between two cases of code design --- code discovery, in which the initial set of stabilizer generators is empty, and code extension, in which the initial set of stabilizer generators is non-empty. Both paradigms are addressed by our code design framework (Sec.~\ref{sec:sat}), and notably both paradigms can be made hardware-aware by incorporating error characterizations. Code extension can generate physics-informed codes by populating the initial set of stabilizer generators with problem symmetries. Framed purely in terms of error correction, code extension can generate new codes from old codes that can detect more errors.

\subsection{Discovery of Quantum Codes} \label{sec:discovery}

The problem of code discovery is to find some set of stabilizer generators constituting a quantum code encoding $k$ logical qubits in $n$ physical qubits that is able to detect or correct some desired set of Pauli errors.  
Let 
\begin{equation}
    \mathcal{E} = \{E_1, \ldots, E_s\}
\end{equation}
be the set of (Pauli) errors we would like to be correctable. Let $C$ be the proposed code encoding $k$ logical qubits using $n$ physical qubits with stabilizer group $\S$ as in Eq.~\eqref{eqn:stabilizer-code-generators}. Then $C$ detects the errors in $\mathcal{E}$ if, for any $E \in \mathcal{E}$, either $E$ anticommutes with at least one stabilizer generator $g_i$ and is thus detectable, or $E \in \S$ such that it does not corrupt the encoded information (\textit{i.e.}, $E$ is a logical identity). For $C$ to be an error-correcting code for the set of errors $\mathcal{E}$, the stabilizer group must fulfill the Knill-Laflamme conditions~\cite{Knill_Laflamme_1996,nielsen2000}; for all pairs of errors $E_i, E_j \in \mathcal{E}$, it must be true that
\begin{equation}
    \label{eq:knill-laflamme}
    E_i E_j^\dagger \notin \mathcal{N}(\S) \setminus \S\, ,
\end{equation}
where $\mathcal{N}(\S)$ is the normalizer of $\S$.
Generating the normalizer $\mathcal{N}(\S)$ can be done in polynomial time with Gaussian elimination --- we give an explicit algorithm for this in Appendix~\ref{app:normalizer}.

\subsection{Extensions of Quantum Codes} \label{sec:extension}

Instead of discovering a new code every time one wishes to correct more errors in encoded information, it is appealing to modify an existing code in such a way that the new code has a higher distance. In classical error correction, one such modification is code \textit{extension}, where $m$ bits are added to the end of the codewords of an $[n,k,d]$ classical error-correcting code ~\cite{hill_extension_1999, hill_extensions_1995, grassl_computing_2007}. If the new bits are chosen judiciously, the extended code can correct more errors. Previous work on extending quantum codes used techniques from machine learning~\cite{olle_simultaneous_2024,yanay2026learningbettererrorcorrection} and concatenated codes~\cite{cao2025growingsparsequantumcodes,PhysRevResearch.7.023086}, but these works do not employ the generalized formalism presented here.

We begin by reviewing the concept of classical code extensions as introduced in Ref.~\cite{grassl_computing_2007}. Let $C$ be an $[n,k,d]_q$ classical code over $\mathbb{F}_q$ with minimum distance $d$. Let $G \in \mathbb{F}_q^{n \times k}$ be a generator matrix for $C$ of full rank.
(Note that we set codewords to be columns of the generator matrix whereas Ref.~\cite{grassl_computing_2007} sets codewords to be rows.)
By $\mathcal{S}_d = \{c \in C : |c| = d\}$ we denote the set of all codewords of Hamming weight $d$ and by $\mathcal{I}_d = \{v \in \mathbb{F}_q^k : ~|Gv|=d\}$ we denote the set of corresponding information vectors, i.e., the unencoded vectors $v \in \Ftwo^k$ that generate weight-$d$ codewords. Then, Ref.~\cite{grassl_computing_2007} shows that the code $C$ can be extended to a code $C'$ with parameters $[n+m, k, d+1]_q$ if and only if there exists a matrix $X \in \mathbb{F}_q^{m \times k}$ such that
\begin{equation}
    \sum_{i=1}^k X_i v_i \neq 0,\quad \forall v \in \mathcal{I}_d,
\end{equation}
where $X_i$ denotes the $i$\textsuperscript{th} column of $X$. The extended code $C'$ has generator matrix
\begin{equation} \label{eqn:generator-matrix-classical-extension}
    G' =
    \begin{bmatrix}
        G \\
        X \\
    \end{bmatrix}
    .
\end{equation}

Here, we generalize this notion to quantum code extension. For this, it is helpful to rewrite Eq.~\eqref{eqn:stabilizer-code-generators} in terms of the check matrix.

\begin{lemma}[Check Matrix for Classical Code Extension]\label{th:classic-ext} 
    Let $C$ be an $[n,k,d]$ classical code with check matrix $H \in \Ftwo^{(n-k)\times n}$. Let the code $C' = [n+m, k, d+1]$ be an extension of $C$.
    Then the check matrix $H'$ of the extended code $C'$ can be expressed in the form
    \begin{equation} \label{eqn:check-matrix-classical-code-extension}
      H' =
      \begin{bmatrix}
          H   & \mathbb{0}_{(n-k)\times m} \\
          U_1 & U_2
      \end{bmatrix}
  \end{equation}
    where $U_1 \in \Ftwo^{m\times n}$ and $U_2 \in \Ftwo^{m\times m}$ together form
    the $m$ extra checks $U = [U_1 \mid U_2]$. The original checks are unchanged,
    with zeros on the $m$ new bits.
\end{lemma}

\begin{proof}
    Let $G'$ be the generator matrix of the extended code as in Eq.~\eqref{eqn:generator-matrix-classical-extension}. To form a valid check matrix for $C'$, the rows of the check matrix $H'$ must be orthogonal to the columns of $G'$. Let $h_i$ be the $i$\textsuperscript{th} row of $H$, and construct $h'_i = [h_i\ \mathbb{0}_{1\times m}]$ from $h_i$. Automatically, $h_i$ is orthogonal to any column $g_j$ of $G$. For any column $g_j$ of $G$, the corresponding column of $G'$ has the form $g'_j = [g_j\ x_j]$, where $x_j$ is the $j$\textsuperscript{th} column of $X$. The inner product of $h'_i$ and $g'_j$ is
    \begin{equation}
        h'_i \cdot g'_j = \bigoplus_{l=1}^{n+m} h'_{i[l]} g'_{j[l]} = (h_i \cdot g_j) \oplus (\mathbb{0}_{1\times m} \cdot x_j) = 0.
    \end{equation}
    In this way, $n-k$ vectors orthogonal to the columns of $G'$ have been constructed. The new check matrix $H'$ has $n+m-k$ rows. To complete the check matrix, $m$ more vectors orthogonal to the columns of $G'$ are found and appended to $H'$. Let these new rows form the matrix $U$. Then the matrix $H'$ defined in Eq.~\eqref{eqn:check-matrix-classical-code-extension}
    has $n+m-k$ rows that are all orthogonal to the columns of $G'$, and thus is a valid check matrix for the code $C'$.
\end{proof}

This allows us to generalize code extension to quantum codes. 

\begin{definition}[Quantum Code Extension] \label{def:quantum-code-extension}
    Let $C$ be a quantum code with parameters $[\![n,k,d]\!]$ and a parity check matrix $H=[H_X|H_Z]$. A code $C'$ with parameters $[\![n+m,k',d']\!]$ is an extension of $C$ if and only if it admits a parity check matrix of the form
    \begin{equation}
        H'=
        \begin{bNiceArray}{cc|cc}
            H_X & W_X & H_Z & W_Z \\
            \multicolumn{2}{c}{\text{---}\; U_X \;\text{---}}
            &
            \multicolumn{2}{c}{\text{---}\; U_Z \;\text{---}}
        \end{bNiceArray}
        .
    \end{equation}
    where $W_X, W_Z \in \Ftwo^{(n-k) \times m}$ and $U_X, U_Z \in \Ftwo^{m \times (n+m)}$ are matrices that represent new stabilizers added to the code $C$ to form $C'$.
\end{definition}

Note that while the classical code extension Eq.~\eqref{eqn:check-matrix-classical-code-extension} leads to the original checks being $0$ on the new bits added during the extension, we do not make such a restriction for quantum code extensions. Instead, we simply require that the upper-left block of the check matrix remains unchanged in the new code and keep $W_X, W_Z \in \Ftwo^{(n-k) \times m}$ arbitrary. In the case that $W_X, W_Z = \mathbb{0}_{(n-k) \times m}$ we refer to the extension as an \textit{extension without mutation}, and in the case that $W_X, W_Z \neq \mathbb{0}_{(n-k) \times m}$ we refer to the extension as an \textit{extension with mutation}.

As an example, consider a three-qubit repetition code with stabilizer generators $g_1 = ZZI$ and $g_2=IZZ$. The code defined by these stabilizer generators is not able to detect any phase flip errors, since they commute with the stabilizer generators. In order to detect these errors, an extension can be employed that adds one qubit (\textit{i.e.} $m=1$) and one stabilizer generator $g_3 = \bigotimes_{i=1}^4 \sigma_i$, where $\sigma_i$ is a single-qubit Pauli operator. To detect weight-one phase flip errors on the original three qubits of the code, $g_3$ must be made up of Pauli-$X$ operators. In order to keep the stabilizer group abelian, choose $\sigma_1 = \sigma_2 = \sigma_3 = X$. Because the code has three stabilizer generators and encodes one logical qubit, we must have $n=4$ for the extended code. At least one stabilizer must have non-trivial support on the fourth physical qubit, and the original stabilizer generators from the repetition code do not have this feature, so it is required that $\sigma_4 \neq I$. For simplicity, choose $\sigma_4 = X$. The new stabilizer $g_3 = X^{\otimes 4}$ enables the detection of weight-one phase flip errors on each of the qubits, and the code with stabilizer group $\langle g_1, g_2, g_3 \rangle$ can still detect the bit flip errors on the first three qubits.

Since the purpose of extending a classical code is to increase the number of errors that it can correct, it is natural to ask this question for quantum code extensions. This is not immediately obvious because, although the new checks of the extended code are designed to detect higher-weight errors, in order to prevent the distance from decreasing, the new checks must also be able to detect errors on the new qubits at the same time. We show below that this is indeed possible.

\begin{theorem}[Existence of Distance-Increasing Code Extensions]\label{thm:existence-of-distance-increasing-extensions}
    Given an $[\![n,k,d]\!]$ code $C$ with $d > 1$, there exists an extension $C'$ of $C$ with parameters $[\![n+m,k',d']\!]$  where  $1 \leq m \leq n(n-1)$ with $d'>d$ and $k' \geq k$.
\end{theorem}

\begin{proof}
    Let $C'$ be the code that consists of $C$ concatenated with itself. It is known~\cite[Chp.~3.5]{Gottesman} that $C'$ is an $[\![n^2,k^2,d^2]\!]$ code. Furthermore, all stabilizers of $C$ are also stabilizers of $C'$. (Each inner block of $C'$ is a copy of $C$.) Since $d > 1$, it follows that $d' = d^2 > d$, and similarly $k' = k^2 \ge k$.  So, letting $m=n^2 - n = n(n-1)$, we see that $C'$ is an extension of $C$ with parameters $[\![n+m,k',d']\!]$ with $d'>d$ and $k' \ge k$.
\end{proof}

\begin{figure}
    \centering 
    \includegraphics[width=0.49\linewidth]{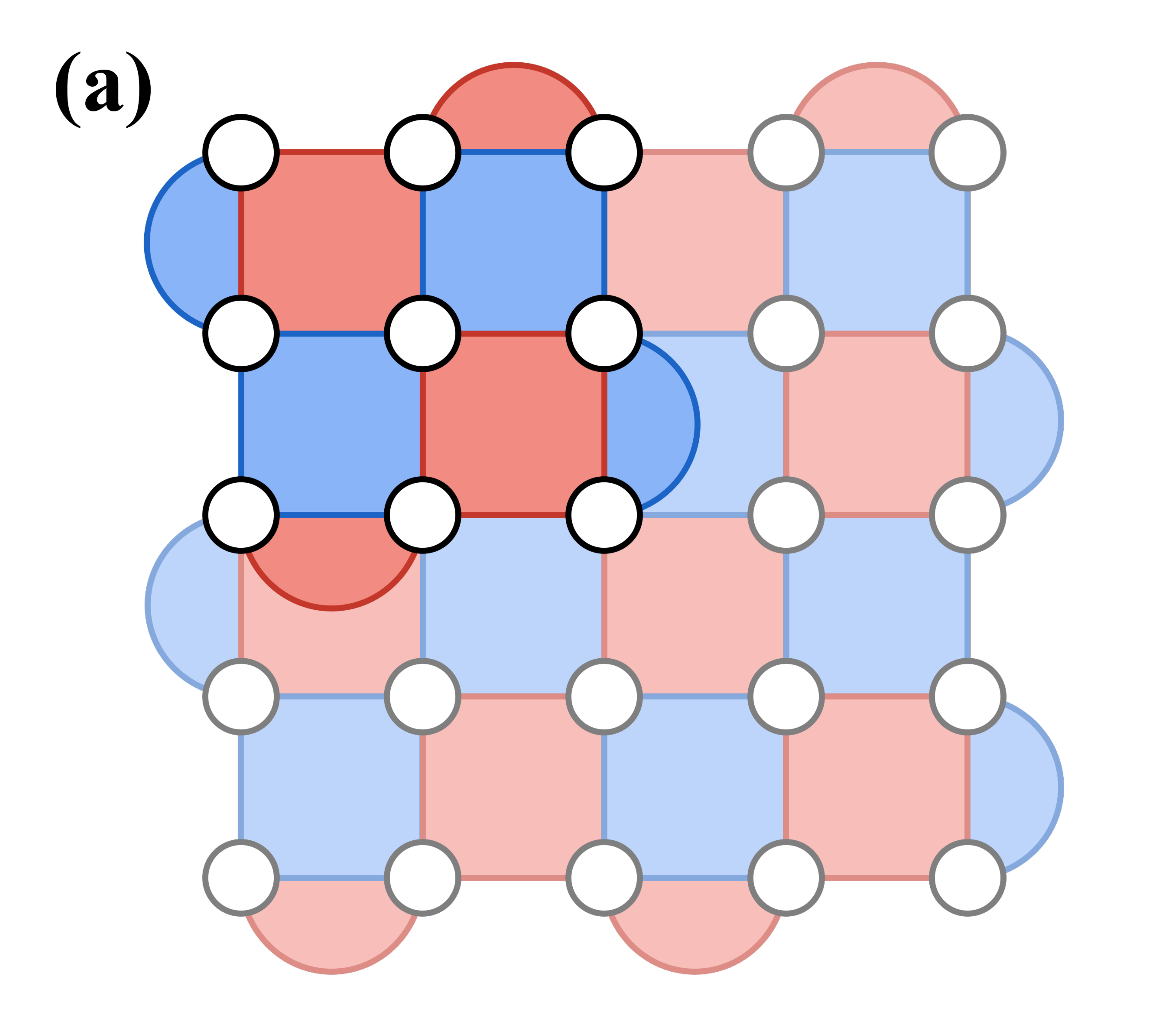} 
    \includegraphics[width=0.49\linewidth]{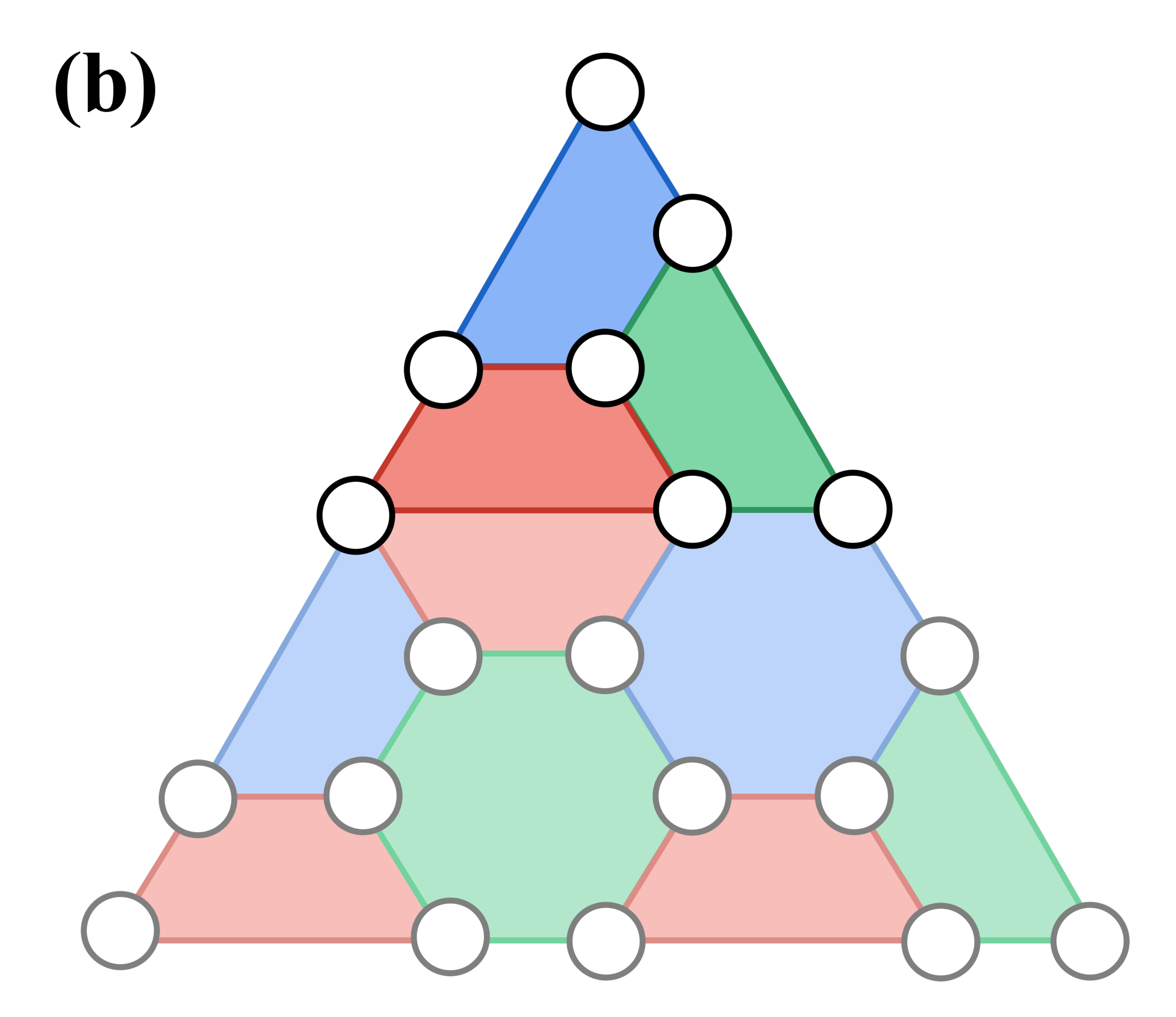}
    \caption{Illustration of distance-increasing extensions of quantum codes with mutation, shown here for two QLDPC codes. (a) A distance $d=3$ rotated surface code  overlaid on a $d=5$ distance surface code. (b) A distance $d = 3$ triangular color code (the Steane code) overlaid on a distance $d = 5$ triangular color code. Note that, in both cases, there are checks along the boundary of the smaller code that must be altered. Thus, these are examples of distance-increasing extensions with mutation.}
    \label{fig:qldpc-growth}
\end{figure}

Our proof here relies on code concatenation, which can be both resource intensive (in the number of logical qubits) as well as difficult to decode relative to Calderbank-Shor-Steane (CSS) codes. A lower overhead route is through quantum low density parity check (QLDPC) codes such as surface codes~\cite{acharya_quantum_2025}, color codes~\cite{lacroix_scaling_2025}, and bivariate bicycle (BB) codes~\cite{wang_demonstration_2026}. These code families are defined on a lattice, and increasing the lattice size (instead of concatenation) is used to increase their distance. For example, Fig.~\ref{fig:qldpc-growth}(a) shows rotated surface codes with $d=3$ and $d=5$. The nine physical qubits of the $d=3$ code are associated with those in the upper-left corner of the larger $d=5$ code. Every four-body check of the $d=3$ code is also a four-body check of the $d=5$ code. The two-body checks along the top and left edges of the $d=3$ code's lattice are present in the $d=5$ code as well. However, the two-body checks along the right and bottom edges correspond to four-body checks in the $d=5$ code. We say that the $d=5$ code is an extension \textit{with mutation} of the $d=3$ code. The same is true in Fig.~\ref{fig:qldpc-growth}(b) for triangular color codes with distance $d=3$ (the Steane code) and $d=5$. Associating the qubits of the $d=3$ code with the upper part of the triangle of the $d=5$ code, these two codes are also related by an extension. Mutation is again necessary, since a trapezoidal face in the $d=3$ code must be turned into a hexagon in the $d=5$ code. Interestingly, bivariate bicycle codes are a class of QLDPC codes which admit extension without mutation.
\begin{theorem}
    \label{thm:bb-extension}
    Let $C$ be the bivariate bicycle code defined by the quadruple $(\ell, m, A, B)$. Let $A'(x,y) = A(x, y^t)$ and $B'(x,y) = B(x,y^t)$ for some integer $t > 1$. The bivariate bicycle code $C'$ defined by $(\ell, tm, A', B')$ is an extension of $C$ without mutation up to a permutation of the physical qubits in $C'$.
\end{theorem}
\noindent A proof of Theorem~\ref{thm:bb-extension}, including background on bivariate bicycle codes, is given in Appendix~\ref{sec:bb-proof}.

Beyond these examples, there is no general framework for producing extensions of quantum codes (up to the discussion of prior work in Sec.~\ref{sec:intro}, but these papers deal with code discovery and not code extension). In this paper, we provide such a framework, and discover interesting classes of quantum codes that are both physics-informed and hardware-aware, having certain advantages with respect to state-of-the-art quantum codes. As our framework is computational, it is crucial to study the complexity of the code design problem, which is done in the following section.

\section{Complexity of Quantum Code Design} \label{sec:complexity}

Previously-studied approaches to designing quantum codes often employ heuristic search through the set of possible check matrices \cite{bravyi_high-threshold_2024, iv_low-weight_2024} or the solution to difficult optimization problems such as game-theoretic methods \cite{guerrero_game-theoretic_2026}, maximum clique cover \cite{chuang_codeword_2009}, or the use of reinforcement learning to optimize a cost function \cite{olle_simultaneous_2024}. In Ref.~\cite{peham_automated_2025}, it is shown that the problem of designing a quantum code bears a strong resemblance to a set cover problem, an NP-complete problem. It has been shown that finding the minimum distance of a code (though not necessarily designing a code with a certain distance) was previously shown to be NP-Hard~\cite{kapshikar2023}. In the realm of classical error correction, SAT has been proposed as a design strategy \cite{shamshiri_error-locality-aware_2010}, and solutions to systems of Boolean polynomial equations have been used to compute extension of classical codes~\cite{grassl_computing_2007}. 
However, to our knowledge, no work to date has rigorously analyzed the computational complexity of \textit{designing} quantum codes. While mappings \textit{from} code design to an NP-complete problem are widely used, a mapping from an NP-complete problem \textit{to} a code design problem (and thus a proof of complexity) has not been demonstrated in prior work. Such a proof is necessary to rule out the existence of more efficient algorithms. In this section, we prove that the code design problem is NP-complete. 

Before analyzing the time complexity of the problem, we outline the memory resources required to find a solution. When discovering a quantum code, every element of the check matrix $H \in \Ftwo^{(n-k) \times 2n}$ is unknown, defining a $2n(n-k)$-dimensional search space. The matrix $H$ is constrained by the facts that the resulting stabilizer group must commute and that some desired set of errors must be detectable (or, equivalently, correctable). The problem of code extension outlined in Sec.~\ref{sec:extension} is solved by finding values of the free variables $U$ and $W$ given some initial stabilizers and a set of errors. The free variables must be chosen such that the whole stabilizer group is abelian and all desired errors, both those that were previously detectable and those that were undetectable, can be detected by the new code. Given that $U_X, U_Y \in \Ftwo^{m \times n+m}$ and $W_X, W_Y \in \Ftwo^{(n-k)\times m}$, the number of variables in the problem scales as $O((n-k)m + nm^2)$. The source of the worst spatial scaling is storing the Pauli errors that the code must correct. If one desires to discover a distance $d$ code, then all errors of weight less than $d$ must be enumerated. The total number of errors is
\begin{equation}
    M = \sum_{w=1}^{d-1} 3^w {n+m \choose w},
\end{equation}
where $n+m \choose w$ counts the number of ways to choose the $w$ qubits comprising the support of a weight-$w$ Pauli error. For each such support, each of the $w$ qubits can independently be assigned to one of the three non-identity Pauli operators, resulting in $3^w$ distinct Pauli errors.

Discovering a high-distance code or extending a code to detect a large set of high-weight errors is thus memory-intensive. 
These memory scaling concerns are mitigated if we restrict to discovering low-distance codes or only correcting a few additional errors in a code extension problem.

In order to prove that the code design problem itself is NP-complete, we define the following decision problem version of $k$-\textsc{code design}.

\begin{definition}[$k$-\textsc{code design} (Decision problem)]
    \label{def:code_design}
    Given a set of $n$-qubit Pauli operators $\Sigma \subseteq \mathcal{P}_n$, 
    a set of $n$-qubit Pauli errors $\mathcal{E}$, 
    and an integer $k$, 
    decide whether there exists a subset of at most $n-k$ linearly independent operators from $\Sigma$ 
    that generates the stabilizer group of an $[\![n,k]\!]$ stabilizer code detecting all errors in $\mathcal{E}$, that is, every $E\in\mathcal{E}$ either has a non-zero syndrome or belongs to the stabilizer group.
\end{definition}

To prove the problem is NP-complete, we utilize a reduction from $k$-\textsc{set cover}, which was shown to be NP-complete by Karp~\cite{karp_reducibility_1972}.

\begin{definition}[$k$-\textsc{set cover}]
    Given a finite universe of values $U = \{1, \ldots, r\}$, a set of $n$ finite subsets of $U$, $\{S_j\}_{j=1}^n$, and an integer $k$, the problem $k$-\textsc{set cover} asks to decide if there is a smaller set of at most $k$ subsets $S_{i_1}, \ldots, S_{i_k}$ s.t. $\bigcup_{l=1}^k S_{i_l} = U$.
\end{definition}

Using this, we can show the following:

\begin{theorem} \label{thm:np-complete}
    $k$-\textsc{code design} is NP-complete.
\end{theorem}

\begin{proof}
    To prove that the problem $k$-\textsc{code design} is NP-complete, we will first prove that it is NP-hard by reducing $k$-\textsc{set cover} to $k$-\textsc{code design}, then we prove membership in NP.

    First, we reduce an instance of $k$-\textsc{set cover} to an instance of $k$-\textsc{code design}. Suppose we are given a set of subsets $\{S_j\}_{j=1}^n$ drawn from a universe $U=\{1,\ldots,r\}$ and an integer $k$; we will show that this comprises an instance of $k$-\textsc{code design}. We construct candidate stabilizers $\Sigma = \{\hat{S}_1, \ldots, \hat{S}_n\}$ and errors $E_1, \ldots, E_r$, together with some number of logical qubits $\bar{k}$ and physical qubits $\bar{n}$ to specify an instance of $k$-\textsc{code design} whose solution encodes a solution to our instance of $k$-\textsc{set cover}.
    
    Let a Pauli error $E_i$ be assigned to each element $u_i$ of the universe $U$. Let $\mathcal{E} = \{E_1, \ldots, E_r\}$, and define $E_i = X_i$. Each stabilizer $\hat{S}_j$ is constructed from a subset $S_j$ s.t. $\hat{S}_j$ anticommutes with the error $E_i$ iff $i \in S_j$. One valid choice is $\hat{S}_j = \bigotimes_{i=1}^{\bar{n}} \hat{S}_{j[i]}$, where
    \begin{equation}
        \hat{S}_{j[i]} =
        \begin{cases}
            I & i \notin S_j, \\
            Z & \textrm{else.} \\
        \end{cases}
    \end{equation}
    For all qubits with index $i>r$, the value does not affect the ability to detect the errors $E_i$, so the values on those qubits will be determined later. Since all stabilizers defined this way are in $\{I,Z\}^{\otimes \bar{n}}$, they all commute. Since all errors are in $\{I,X\}^{\otimes \bar{n}}$, the errors cannot be in any subgroup generated by some subset of $\Sigma$. To find a code that encodes $\bar{k}$ qubits and detects all errors in $\mathcal{E}$, it suffices to choose generators $\{\hat{S}_{i_1}, \ldots, \hat{S}_{i_{\bar{n}-\bar{k}}}\} \subset \Sigma$ s.t. for each $E_i \in \mathcal{E}$, there is at least one generator $\hat{S}_{i_l}$ in the subset that anticommutes with $E_i$.

    Any subset of $l$ of our candidate stabilizers $\Sigma$ should be independent so that they can generate a group of size $2^l$. If the whole set $\Sigma$ is independent, then any subset of $\Sigma$ will also be independent. Taking a value of $\bar{n} > r$ enables this requirement to be satisfied. Represent each stabilizer generator $\hat{S}_j$ by a binary vector $b_j \in \Ftwo^{\bar{n}}$. For now, $b_{j[i]} = 0$ for $i > r$. Suppose there is some algorithm that could find the linearly independent subset of these binary vectors in polynomial time (analogous to Gram-Schmidt). Then there are potentially some vectors $b_j$ that are not in the independent set. We can change values on the bits $r+1$ through $\bar{n}$ to make them linearly independent. These changes do not affect the ability to detect errors on the first $r$ qubits. Some value of $\bar{n}$ will be needed for this procedure; indeed, only one extra bit is needed. If some vector $b_l$ is in the linearly independent set, set $b_{l[r+1]} = 0$, and $b_{l[r+1]} = 1$ otherwise.

    Finally, we must know how many logical qubits $\bar{k}$ the code must encode. Since we desire $k$ stabilizers in a valid solution to $k$-\textsc{code design}, $k = \bar{n} - \bar{k}$, which implies that $\bar{k} = \bar{n} - k$.
    The errors $E_i$ can be constructed from the universe $U$ in their binary symplectic forms in $O(r \bar{n})$ time. The stabilizers $\hat{S}_j$ can be constructed from the subsets $S_j$ in $O(n \bar{n})$ time.

    Suppose there is an error-detecting code that solves the constructed instance of $k$-\textsc{code design}. This code is mapped onto the solution to the original instance of $k$-\textsc{set cover} in the following way: For each stabilizer $\hat{S}_{i_l}$ in the code, we enumerate the errors $E_{j_1}, \ldots, E_{j_q}$ that $\hat{S}_{i_l}$ anticommutes with. Equivalently, enumerate all qubits with index $l \leq r$ s.t. $\hat{S}_{j[l]} = Z$. The detectable errors for stabilizer $\hat{S}_j$ correspond to a subset $S_j = \{j_1, \ldots, j_q\}$. Because the code detects all errors, there is at least one stabilizer that anticommutes with any given $E_i$, and thus there is at least one corresponding subset that contains the value $i$. The set of the subsets corresponding to the chosen stabilizer generators is a set covering and has size $k$. The existence of an error detecting code for the constructed instance of $k$-\textsc{code design} implies the existence of a set covering for the original instance of $k$-\textsc{set cover}.
    
    Suppose, however, that no code to detect errors $\mathcal{E}$ and encode $\bar{k}$ qubits can be made from the candidate generators $\Sigma$. For any subset $\sigma$ of $\Sigma$ with size $k$, there is at least one error $E_i \in \mathcal{E}$ that commutes with all operators in $\sigma$. Let $\sigma = \{\hat{S}_{j_1}, \ldots, \hat{S}_{j_k}\}$. By construction, $i \notin S_{j_l}$ for all $l = 1, \ldots, q$. Since $i$ is not in the set of subsets corresponding to any set of stabilizers $\sigma$, a set covering does not exist. Thus any instance of $k$-\textsc{set cover} can be reduced to an instance of $k$-\textsc{code design} in polynomial time, meaning $k$-\textsc{code design} is NP-hard.

    Finally, we show that any solution to $k$-\textsc{code design} can be checked in polynomial time. Let $\mathcal{S} = \langle \hat{S}_{j_1}, \ldots, \hat{S}_{j_{\bar{n} - \bar{k}}} \rangle$ be a proposed stabilizer group for the code. For any error $E \in \mathcal{E}$, either there must be a stabilizer $\hat{S}_{j_l}$ that anticommutes with $E$ or $E \in \mathcal{S}$. To check for anticommutation, if $|\mathcal{E}| = r$, then $(\bar{n} - \bar{k})r$ anticommutation relations must be evaluated. Using the binary symplectic form of the Pauli strings, these checks can be done in $O(\bar{n})$ time. Checking whether or not $E \in \mathcal{S}$ can also be done in the binary symplectic form. If $E \in \mathcal{S}$, then $\exists d \in \Ftwo^{\bar{n} - \bar{k}}$ s.t. $E = \hat{S}_{j_1}^{d_1} \cdots \hat{S}_{j_{\bar{n} - \bar{k}}}^{d_{\bar{n}-\bar{k}}}$. Let $E$ have binary symplectic form $\tilde{E}$ and $\hat{S}_{j_l}$ have a binary symplectic form $\tilde{S}_{j_l}$. The equivalent group membership condition is then
    \begin{equation}
        \exists d \in \Ftwo^{\bar{n} - \bar{k}}
        ~\textrm{s.t.}~
        \tilde{E} = \bigoplus_{l=1}^{\bar{n} - \bar{k}} d_l \tilde{S}_{j_l}
    \end{equation}
    by the fact that multiplication in the Pauli subgroup is equivalent to bitwise addition of the binary symplectic forms modulo 2. The problem, then, is to prove that there is a solution to a system of Boolean linear equations with $2 \bar{n}$ equations and $\bar{n} - \bar{k}$ unknowns. This check can be done in polynomial time by rewriting the problem as an instance of XOR-SAT and proving whether the formula is satisfiable~\cite{schaefer_complexity_1978}. Both steps (checking for anticommutation and checking for group membership) take polynomial time, and must be repeated once for each error. A candidate solution to an instance of $k$-\textsc{code design} can thus be checked in polynomial time in the number of physical qubits, logical qubits, and errors. Thus the problem $k$-\textsc{code design} is in NP.
\end{proof}

\section{Quantum Code Design as a Satisfiability Problem} \label{sec:sat}

In this section, we formulate the quantum code design problem as an instance of SAT. This allows us to apply state-of-the-art SAT solvers to design codes, and to analyze the practical runtime of relevant problem instances. 

\subsection{SAT-Based Code Discovery}
Consider the quantum code design problem (Def.~\ref{def:code_design}), such that the goal is to build an $[\![n,k]\!]$ quantum code that detects each error in the set $\mathcal{E} = \{E_1, \ldots, E_s\}$. If the goal is to find an $[\![n,k,d]\!]$ code, then $\mathcal{E} = \bigcup_{w=1}^{d-1} \{P : P \in \mathcal{P}_n, |P|=w\}$, where $|\cdot|$ denotes the weight of a Pauli string. The solution to this problem is a set of stabilizer generators $g_1, \ldots, g_{n-k}$. Let the unknown stabilizer generators have binary symplectic forms $u_1, \ldots, u_{n-k}$ and let the errors have binary symplectic forms $\epsilon_1, \ldots, \epsilon_s$. We can also refer to the parts of the binary vectors that correspond to $X$ or $Z$ Pauli operators: $u_i = [u_i^X | u_i^Z]$ and $\epsilon_j = [\epsilon_j^X | \epsilon_j^Z]$.

The stabilizer group formed by the unknown stabilizers must be abelian. In terms of their operator representations, this condition reads $[g_i, g_j] = 0,\ \forall i, j = 1, \ldots, n-k$. The equivalent constraint on the binary symplectic forms is given by the function
\begin{equation}
    \label{eq:constraint-commutator-new-new}
    \texttt{C}_\texttt{U}(u) = \bigwedge_{i=1, j > i}^{n-k} \neg \langle u_i, u_j \rangle.
\end{equation}
If the stabilizer generators mutually commute for some values $u_1, \ldots, u_{n-k}$, then $\texttt{C}_\texttt{U}(u) = 1$; otherwise $\texttt{C}_\texttt{U}(u) = 0$.

In order to detect all errors in $\mathcal{E}$, for every error $E \in \mathcal{E}$, it must be true that either $E$ anticommutes with at least one stabilizer generator of the code or $E$ is in the stabilizer group. A syndrome bit corresponding to generators $g_i$ can be evaluated as $\langle u_i, \epsilon_j \rangle$. To assert that the syndrome is not all zero, we take the logical disjunction of all syndrome bits
\begin{equation}
    \label{eq:constraint-syndrome-not-zero}
    \texttt{S}_j(u) = \bigvee_{i=1}^{n-k} \langle u_i, \epsilon_j \rangle.
\end{equation}
Alternatively, we can check if the operator $E_j$ is in the stabilizer group $\S$. If an error $E \in \S$ up to a phase, then $\exists s \in \Ftwo^{n-k}$ s.t. $E = g_1^{s_1} \cdots g_{n-k}^{s_{n-k}}$. In terms of their binary symplectic forms, this expression becomes
\begin{equation}
    \label{eq:membership-equality-form}
     \epsilon_j = s_{1} u_1 \oplus \cdots \oplus s_{n-k} u_{n-k}.
\end{equation}
Using the fact that $a \oplus b = 1$ if $a \neq b$ and $a \oplus b = 0$ if $a = b$, Eq.~\eqref{eq:membership-equality-form} can be rewritten as
\begin{equation}
    \label{eq:membership-xor-form}
     \epsilon_j \oplus s_{1} u_1 \oplus \cdots \oplus s_{n-k} u_{n-k} = \vec{0},
\end{equation}
where $\vec{0}$ is the vector of zeros with $2n$ elements. In the SAT formulation, the vector $s$ is treated as a set of slack variables. The Boolean function that checks for group membership conditioned on the slack variables is
\begin{equation}
    \label{eq:constraint-in-group}
    \texttt{G}_j(u; s) = \bigwedge_{i=1}^{2n} \neg(\epsilon_j \oplus 
    s_{1} u_1 \oplus \cdots \oplus s_{n-k} u_{n-k})_{[i]},
\end{equation}
where the notation $[i]$ refers to the $i$\textsuperscript{th} element of a binary vector. If the error $E_j$ is in the stabilizer group, then the above formula is satisfiable for some assignment of the variables $s$. Otherwise, the formula is unsatisfiable. When Eqs.~\eqref{eq:constraint-syndrome-not-zero} and \eqref{eq:constraint-in-group} are combined, we obtain the condition requiring all errors in the set $\{\epsilon_j\}_{j=1}^s$ to be detectable
\begin{equation}
    \label{eq:constraint-error-detectable}
    \texttt{D}(u; s) = \bigwedge_{j=1}^s (\texttt{S}_j(u) \vee \texttt{G}_j(u; s)).
\end{equation}
The stabilizer group forms a valid code to detect all errors in $\mathcal{E}$ given that
\begin{equation}
    \label{eq:constraint-errors}
    \texttt{D}(u; s) \wedge \texttt{C}_\texttt{U}(u).
\end{equation}
evaluates to true.

\subsection{SAT-Based Code Extension}
We now consider the problem of extending a code to detect errors in $\mathcal{E}$, some of which may already be detected by the original code. Let $h_1, \ldots, h_{n-k}$ denote the binary symplectic forms of the stabilizers of the original code and $u_1, \ldots, u_m$ denote the $m$ unknown stabilizers. The final stabilizer group, which is made up of the original and unknown stabilizers, must be abelian. As in the code discovery problem, the constraint that the unknown stabilizers commute is represented by Eq.~\eqref{eq:constraint-commutator-new-new}. The constraint that the new stabilizer commute with the existing stabilizers is given by the function.
\begin{equation}
    \label{eq:constraint-commutator-old-new}
    \texttt{C}_\texttt{H}(u) = \bigwedge_{i=1}^{n-k} \bigwedge_{j=1}^m \neg \langle h_i, u_j \rangle.
\end{equation}

The constraints enforcing error detectability have similar functional forms, differing only by minor modifications that account for the partition of the stabilizers into existing and unknown sets. The formula that ensures  the syndrome is nonzero is now given by
\begin{equation}
    \label{eq:constraint-syndrome-not-zero-extend}
    \texttt{S}_j(u) = \bigvee_{i=1}^{n-k} \langle h_i, \epsilon_j \rangle \vee \bigvee_{i=1}^{m} \langle u_i, \epsilon_j \rangle.
\end{equation}
In addition, the check for group membership is updated to
\begin{equation}
    \label{eq:constraint-in-group-extend}
    \begin{split}
    \texttt{G}_j(u; s) = \bigwedge_{i=1}^{2(n+m)}
    \neg(&\epsilon_j \oplus  s_{1} h_1 \oplus \cdots \oplus s_{n-k} h_{n-k} \oplus \\
    & s_{n-k+1} u_1 \oplus \ldots \oplus s_{n-k+m} u_m)_{[i]}.
    \end{split}
\end{equation}
We then impose the constraint in Eq.~\eqref{eq:constraint-error-detectable}, with the updated forms of the functions $\texttt{S}_j(u)$ and $\texttt{G}_j(u;s)$, yielding a new form of the function $\texttt{D}(u;s)$. All together, the total constraint for the code extension problem is
\begin{equation}
    \label{eq:constraint-errors-extend}
    \texttt{D}(u; s) \wedge \texttt{C}_\texttt{U}(u) \wedge \texttt{C}_\texttt{H}(u).
\end{equation}

To support extension with mutation, we treat the existing stabilizer generators as Pauli strings where some of the entries --- namely the $X$ and $Z$ bits for the original $n$ qubits --- are fixed and the rest are left variable. The equations for the SAT problem remain mostly unchanged, except for the omission of Eq.~\eqref{eq:constraint-commutator-old-new}, which is now handled by Eq.~\eqref{eq:constraint-commutator-new-new}.

\subsection{From Detection to Correction}
To correct errors instead of detecting them, the stabilizer group must obey the Knill-Laflamme conditions (Eq.~\eqref{eq:knill-laflamme}). For the case where all errors are Pauli strings, the errors are Hermitian and the condition can be rewritten as
\begin{equation}
    \label{eq:knill-laflamme-revised}
    E_i E_j \notin \mathcal{N}(\S) \setminus \S,\quad \forall E_i, E_j \in
      \mathcal{E}.
\end{equation}
The conditions to detect errors in the code discovery or extension problems (Eqs.~\eqref{eq:constraint-errors} and \eqref{eq:constraint-errors-extend}, respectively) could equivalently be expressed as 
\begin{equation}
    \label{eq:detection-condition}
    E_i \notin \mathcal{N}(\S) \setminus \S,\quad \forall E_i \in \mathcal{E}.
\end{equation}
Note that Eq.~\eqref{eq:knill-laflamme-revised} has much of the same structure as Eq.~\eqref{eq:detection-condition}; the difference being that the Knill-Laflamme conditions apply to pairs of errors instead of single errors. In the binary symplectic representation, $E_i E_j \cong \epsilon_i \oplus \epsilon_j$. Substituting $\epsilon_i \rightarrow \epsilon_i \epsilon_j$ in all constraint functions, the updated constraints for code discovery become
\begin{eqnarray}
    \label{eq:constraint-syndrome-not-zero-correction}
    \texttt{S}_{i,j}(u) &=& \bigvee_{l=1}^{n-k} \langle u_l, \epsilon_i \oplus \epsilon_j \rangle,\\
% \end{equation}
% \begin{equation}
    \label{eq:constraint-in-group-correction}
    \texttt{G}_{i,j}(u; s) &=& \bigwedge_{l=1}^{2n} \neg(\epsilon_i \oplus \epsilon_j \oplus 
    s_{1} u_1 \oplus \cdots \oplus s_{n-k} u_{n-k})_{[l]},\nonumber\\
    &&\\
% \end{equation}
% \begin{equation}
    \label{eq:constraint-error-correctable}
    \texttt{D}(u; s) &=& \bigwedge_{i=1}^{s} \bigwedge_{j=i}^{s} (\texttt{S}_{i,j}(u) \vee \texttt{G}_{i,j}(u; s)).
\end{eqnarray}
Comparatively, the conditions for a code extension problem become
\begin{eqnarray}
    \label{eq:constraint-syndrome-not-zero-extend-correct}
    \texttt{S}_{i,j}(u) &=& \bigvee_{l=1}^{n-k} \langle h_l, \epsilon_i \oplus \epsilon_j \rangle \vee \bigvee_{l=1}^{m} \langle u_l, \epsilon_i \oplus \epsilon_j \rangle,\\
    \texttt{G}_{i,j}(u; s) &=& \bigwedge_{l=1}^{2n}
    \neg(\epsilon_i \oplus \epsilon_j \oplus  s_{1} h_1 \oplus \cdots \oplus s_{n-k} h_{n-k} \oplus \nonumber\\
    &&\quad\quad\, s_{n-k+1} u_1 \oplus \ldots \oplus s_{n-k+m} u_m)_{[l]}, 
\label{eq:constraint-in-group-extend-correct} \\
    \label{eq:constraint-error-correctable-extension}
    \texttt{D}(u; s) &=& \bigwedge_{i=1}^{s} \bigwedge_{j=i}^{s} (\texttt{S}_{i,j}(u) \vee \texttt{G}_{i,j}(u; s)).
\end{eqnarray}

\section{Phase Transitions in the \textsc{k-Code Design} Problem\label{sec:phase_transition}}

The reduction developed in Sec.~\ref{sec:complexity} establishes that $k$-\textsc{Code Design} is NP-complete by polynomial-time reduction from \textsc{set cover}, while Sec.~\ref{sec:sat} provides a reduction to  mixed 2- and 3-\textsc{SAT} to readily compute a solution. Beyond worst-case complexity, however, it is well-known that NP-complete problems exhibit rich average-case behavior. 
One of the most striking phenomena is the existence of the \emph{satisfiability phase transition}, where random instances abruptly change from almost always satisfiable to almost always unsatisfiable as an order parameter is varied. 
This transition has been extensively studied for random $k$-SAT and is closely connected to the empirical hardness of complete SAT solvers~\cite{kirkpatrick_1994, gent_1994, Monasson1999, fernandez_de_la_vega_2001, friedrich_2017}. 
Here, we demonstrate that the same phenomenon appears in random instances of $k$-\textsc{Code Design}.

For random \textsc{$k$-SAT}, one considers formulas with $N$ Boolean variables and $M$ clauses. The natural order parameter is then the \emph{clause density}
\begin{align}
    \alpha = \frac{M}{N},
\end{align}
which fixes the number of constraints per variable. 
For fixed $N$, many random formulas are generated at a given $\alpha$, and the satisfiability probability
\begin{align}
    P_{\mathrm{SAT}}
    =
    \frac{\text{\# satisfiable formulas}}
    {\text{\# total formulas}}
\end{align}
is measured. 
At small clause density, nearly every instance is satisfiable, while at sufficiently large density almost every instance is unsatisfiable. 
The critical clause density, $\alpha_c$, at which the crossover occurs is known as the SAT--UNSAT transition~\cite{Monasson1999, silvio_parisi_2017}.

This transition is significant not only because it marks the onset of unsatisfiability, but also because it coincides with the region of greatest computational difficulty. Complete SAT solvers exhibit their largest runtimes near $\alpha_c$, while formulas that are under- or over-constrained are typically solved quickly.
For under-constrained instances, satisfying assignments are abundant and are found with little search. Conversely, for heavily constrained instances, contradictions are rapidly identified, allowing unsatisfiability to be certified efficiently. 
Near the transition, however, satisfying assignments become exponentially rare, forcing the solver to explore a large fraction of the search tree before reaching a conclusion. 
Consequently, the phase transition provides an experimentally observable signature of computational hardness in the average-case behavior.

In general, the hardest instances of NP-hard problems occur at the phase transition~\cite{ZHANG1996223, HOGG19961, coppersmith_2004, Santra_2014}.
We therefore investigate whether an analogous transition occurs for random
instances of $k$-\textsc{Code Design}.
Consider a problem instance consisting of $n$ physical qubits, the candidate
set $\Sigma=\mathcal{P}_n$, and an error set
\begin{align}
    \mathcal{E}=\{E_1,E_2,\ldots,E_m\}.
\end{align}
Following Sec.~\ref{sec:sat}, this decision problem is reduced to an instance of mixed \textsc{2-3-SAT} and solved using a heuristic SAT solver.
To study the average-case complexity, we generate random error sets $\mathcal{E}$ by sampling $m$ Pauli errors from the pool of all nontrivial errors of weight at most $w$, whose size is
\begin{align}
    M_{\leq w}(n) = \sum_{j=1}^{w}3^j {n \choose j}. 
    \label{eq:error_pool_size}
\end{align}
For each $(n,k,w,m)$, many independent instances are generated and the
satisfiability probability $P_{\mathrm{SAT}}$ is estimated.
The number of sampled errors $m$ sets the number of constraints imposed
on the candidate code and therefore provides a natural control parameter for
the SAT--UNSAT transition.
For visualization, it is useful to normalize $m$ by the number of stabilizer
generators $n-k$, which measures the number of detected errors per stabilizer.
For fixed $(n,k,w)$, small $m/(n-k)$ corresponds to weakly constrained instances
that are typically satisfiable, whereas sufficiently large $m/(n-k)$ produces
instances that are typically unsatisfiable.
Although the constraint density $m/(n-k)$ is analogous to the clause density in random $k$-\textsc{SAT}, with both quantities characterizing the balance between available degrees of freedom and imposed constraints, we do not observe a universal critical value of $m/(n-k)$.
Rather, the location of the transition depends on the code parameters
$(n,k,w)$, motivating the parameter-dependent critical
$m_c(n,k,w)$ developed below.

\begin{figure}[tb]
    \centering
    \includegraphics[width=\linewidth]{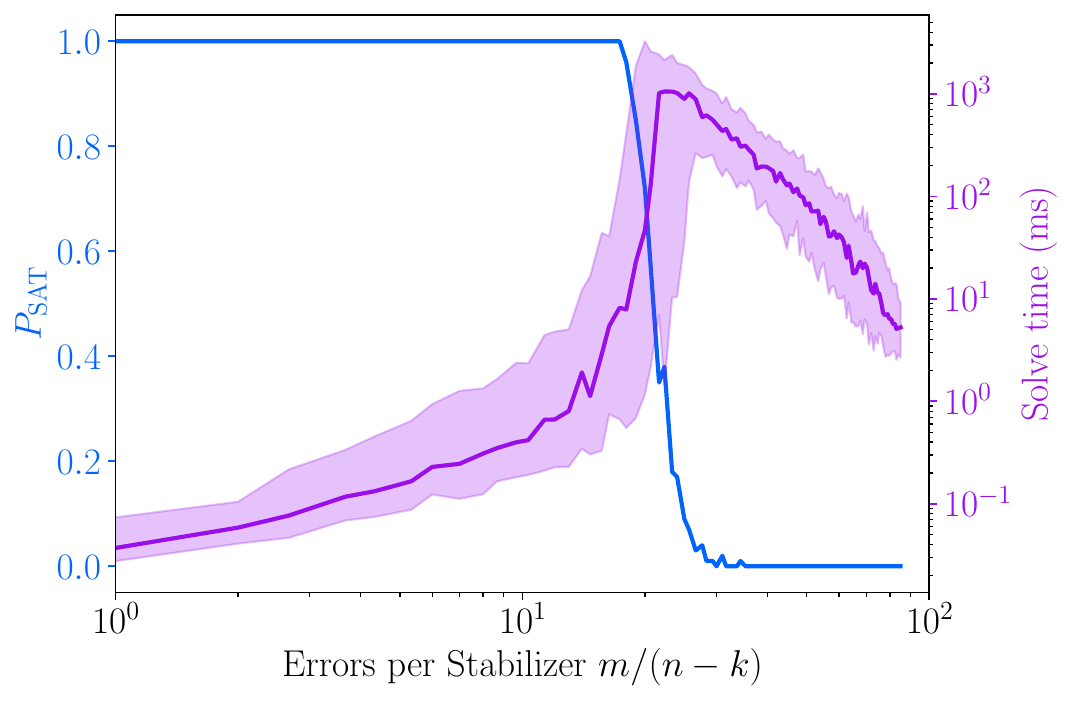}
    \caption{Demonstration of the SAT--UNSAT phase transition which we empirically observe for random instances of \textsc{k-code design} (Def.~\ref{def:code_design}). The order parameter is the number of errors per stabilizer $m / (n - k)$ in the code design problem. In addition to the transition in satisfiability probability $p_\text{SAT}$ at the critical value, the solve time peaks at this critical value. Additional numerical results and cases are discussed in detail in Appendix~\ref{app:phase_transition}.
    }
    \label{fig:phase_transition}
\end{figure}

Figure~\ref{fig:phase_transition} shows the satisfiability probability and SAT-solver runtime as functions of $m/(n-k)$ for the representative case $(n,k,w)=(4,1,4)$.
For these fixed parameters, we observe a sharp finite-size SAT--UNSAT crossover near $m_c/(n-k) \approx 20$, accompanied by a peak in the solver runtime.
Thus, random instances of $k$-\textsc{Code Design} display the characteristic easy--hard--easy behavior associated with satisfiability transitions. Instances far from the crossover are typically resolved quickly, while instances near the transition require substantially greater computational effort.
The position of this crossover changes with $(n,k,w)$ --- we expand upon this point in Appendix~\ref{app:phase_transition}.
So, the transition is characterized by the critical number of errors $m_c(n,k,w)$ rather than by a constraint density $m/(n-k)$.
The coincidence of the satisfiability threshold and the runtime peak therefore demonstrates that random instances of $k$-\textsc{Code Design} possess the same characteristic easy--hard--easy behavior observed in other NP-complete problems~\cite{Monasson1999}.

We develop a finite-size empirical scaling $\hat m_c(n,k,w)$ to determine the critical number of errors $m_c$ at which the problem transitions from SAT to UNSAT.
The model and its details are given in Appendix~\ref{app:phase_transition}.
We can use the predicted critical value $\hat m_c(n,k,w)$ to predict the feasibility of constructing an $[[n,k,d]]$ stabilizer code.
A distance-$d$ code must detect every nontrivial Pauli error of weight less than $d$, so the corresponding \textsc{$k$-Code Design} instance contains $M_{\leq d-1}(n)$ errors.
The quantum Gilbert--Varshamov (QGV) bound provides a sufficient condition for the existence of an $[[n,k,d]]$ stabilizer code~\cite{ekert_macchiavello_qgv_bound_1996, Gottesman}
\begin{align}
    1 + M_{\leq d-1}(n) \leq 2^{n-k},
    \label{eq:gilbert_varshamov_bound}
\end{align}
while the quantum Singleton bound provides a necessary condition~\cite{ketkar2005nonbinarystabilizercodesfinite, Gottesman}
\begin{align}
    n-k \geq 2(d-1).
    \label{eq:singleton_bound}
\end{align}
Thus, codes satisfying Eq.~\eqref{eq:gilbert_varshamov_bound} are guaranteed to exist, whereas codes violating Eq.~\eqref{eq:singleton_bound} cannot exist.
Between these rigorous bounds lies a broad region in which neither bound determines code existence.
We classify this intermediate region using the empirical model developed in Appendix~\ref{app:phase_transition}.

\begin{figure}[t] %[tbh]
    \centering
    \includegraphics[width=0.9\linewidth]{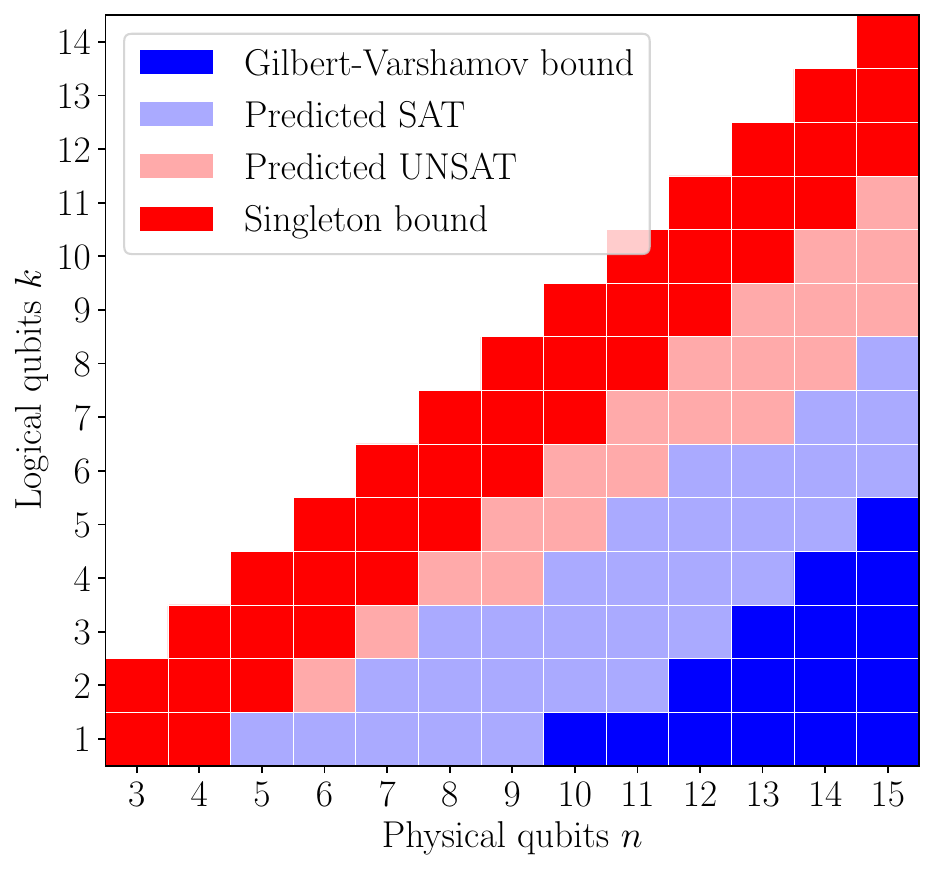}
    \caption{
    Predicted existence of $[\![n,k,3]\!]$ stabilizer codes as a function of the number of physical qubits $n$ and logical qubits $k$.
    Dark blue denotes parameters for which code existence is guaranteed by the quantum Gilbert--Varshamov bound, while dark red denotes parameters excluded by the quantum Singleton bound.
    In the intermediate region, the empirical scaling model for the critical error-set size $\widehat{m}_c(n,k,w)$, with $w=d-1=2$, predicts satisfiable (light blue) and unsatisfiable (light red) \textsc{$k$-Code Design} instances.
    }
    \label{fig:phase_transition_diagram}
\end{figure}

For a target distance $d$, we set $w=d-1$ and compare the complete error-set size $M_{\leq d-1}(n)$ with the predicted critical value $\widehat{m}_c(n,k,d-1)$.
We predict the corresponding \textsc{$k$-Code Design} instance to be satisfiable when
\begin{align}
    M_{\leq d-1}(n) < \widehat{m}_c(n,k,d-1),
\end{align}
and unsatisfiable when the inequality is reversed.
Figure~\ref{fig:phase_transition_diagram} shows the resulting phase diagram for distance-three codes.
The Gilbert-Varshamov and Singleton bounds rigorously constrain the low- and high-rate regions, respectively, while the fitted SAT--UNSAT transition provides an empirical prediction of code existence in the contested region.
The SAT formulation thus provides a rigorous framework to prove the (un)satisfiability of a given set of code parameters. % TODO: A little more conclusion/discussion wanted here.

\section{Physics-Informed Code Design} \label{sec:physics-informed-code-design}

We now demonstrate our framework by designing novel quantum codes starting from a set of problem stabilizers. We refer to this application of code extension as physics-inspired code design since the initial stabilizers arise from physical considerations of the problem. 
This can be seen as a generalization of quantum error detection/symmetry verification~\cite{cai_quantum_2021} as we show how to iteratively build new codes to increase performance as measured by accuracy of computed expectation values. In addition to designing new codes, we study their performance relative to the baseline codes. 

We consider the Fermi-Hubbard model 
\begin{equation}
\begin{aligned}
H ={}&
-t \sum_{\langle i,j\rangle}\sum_{\sigma}
\left(
a^\dagger_{i,\sigma}a_{j,\sigma}
+
a^\dagger_{j,\sigma}a_{i,\sigma}
\right)
+
U\sum_i
a^\dagger_{i,\uparrow}a_{i,\uparrow}
a^\dagger_{i,\downarrow}a_{i,\downarrow}
\\
&-\mu\sum_i\sum_{\sigma}
a^\dagger_{i,\sigma}a_{i,\sigma}
-h\sum_i
\left(
a^\dagger_{i,\uparrow}a_{i,\uparrow}
-
a^\dagger_{i,\downarrow}a_{i,\downarrow}
\right),
\end{aligned}
\end{equation}
where $a_{i,\sigma} (a^\dagger_{i,\sigma})$ is the $i$th site annihilation (creation) operator for spin states $\sigma \in \{ \uparrow, \downarrow \}$.  
The Hamiltonian is further specified by the tunneling amplitude $t$, 
while $U$ is the Coulomb potential, $\mu$ is the chemical potential, and $h$ is the magnetic field. In this model, the numbers of spin-up and spin-down fermions are separately conserved, providing a set of symmetries for the system. Using the Jordan-Wigner encoding and a staggered fermion ordering on the lattice, this conservation of particle number is enforced by a symmetry group with generators
\begin{equation}
    \label{eq:fh-g}
    g_\uparrow = (I \otimes Z)^{\otimes n/2}
    \quad \text{and} \quad
    g_\downarrow = (Z \otimes I)^{\otimes n/2},
\end{equation}
where $n$ is the number of qubits in the system, which is equal to double the number of sites in the lattice. Since they are Pauli strings, these generators can also be thought of as the generators of a quantum error-detecting code. For an $n$-qubit system, the code has parameters $[\![n,n-2,1]\!]$. We call the stabilizer code with generators given by Eq.~\eqref{eq:fh-g} the Fermi-Hubbard code.

\begin{figure}
    \centering
    \includegraphics[width=\linewidth]{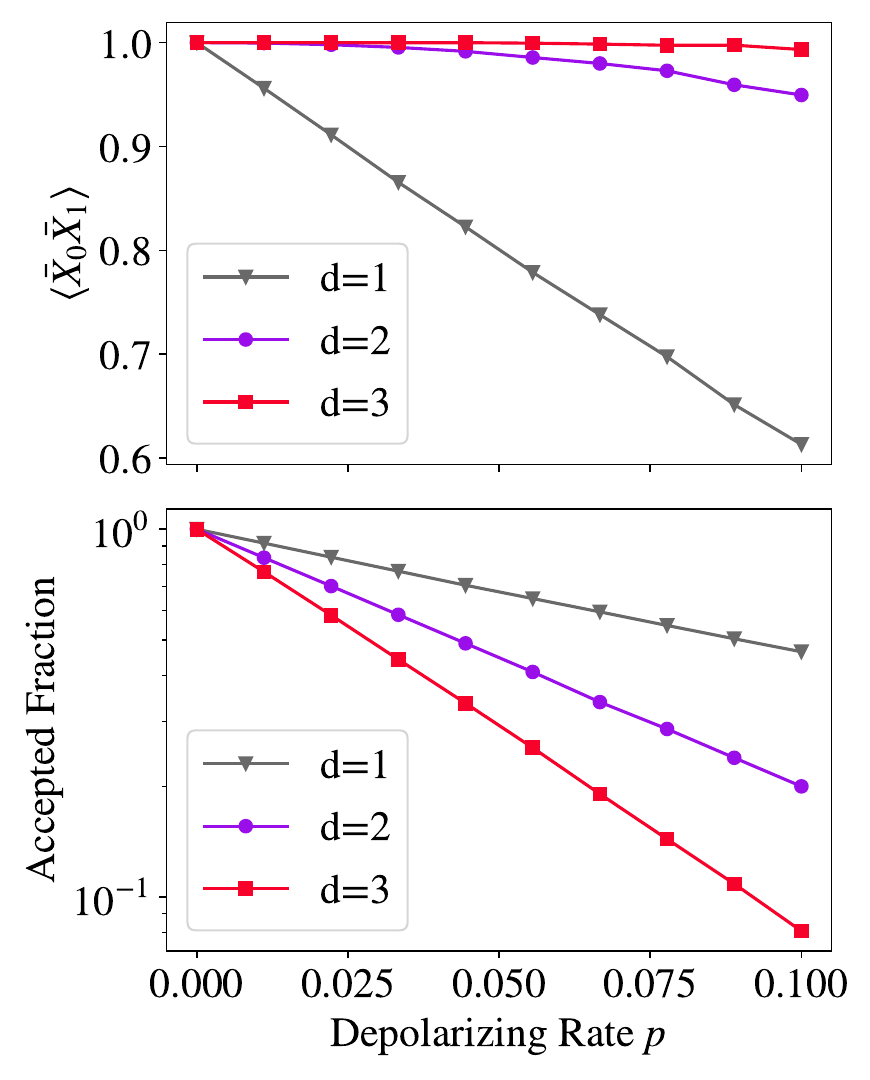}
    \caption{Demonstration of improved accuracy in error-detected expectation values for physics-informed codes. Here we start from the symmetries~\eqref{eq:fh-g} of an $n = 6$ Fermi-Hubbard Hamiltonian and extend to distance $d = 2$ and $d = 3$ codes. (top) Expectation value of the logical observable $\bar{X}_0 \bar{X}_1$ in a logical Bell state vs. depolarizing noise rate $p$. Without noise, this expectation value should be one. Here, logical state preparation is assumed to be done fault-tolerantly, and depolarizing noise acts on physical qubits after each logical operation. As can be seen, extending the codes to higher distances significantly improves the accuracy of the expectation value. As discussed in the main text, this can be viewed as a generalization of symmetry verification~\cite{cai_quantum_2021} to improve error-mitigated expectation values. (bottom) The accepted fraction of samples (i.e., the number of sampled codewords) vs. the depolarizing noise rate $p$. As can be seen, codes extended to higher distances require slightly more samples due to a larger number of physical qubits and therefore larger number of noisy channels. This increase in required shots is significantly less than what would be required for concatenated codes or arbitrary (non physics-inspired) codes. This accepted fraction is present only in error detection where expectation values are computed by filtering in post-processing.}
    \label{fig:fh-error-detection}
\end{figure}

We apply our code design framework to extend the Fermi-Hubbard code and compute the accuracy of error-detected (symmetry verified) expectation values. Fig.~\ref{fig:fh-error-detection} shows the expectation value of the logical observable $\bar{X}_0 \bar{X}_1$ versus the noise rate $p$ (defined below). Here, we start from an $n=6$ Fermi-Hubbard code with distance $d = 1$ and extend to $d = 2$ and $d = 3$. The distance two code has $n=8$ physical qubits and the distance three code has $n=12$ qubits. We simulate an error detection experiment in which the logical state $\ket{0}^{\otimes k}$ is prepared, then logical Hadamard and CNOT operations are applied to prepare a Bell state on the first two logical qubits. In the logical Bell state, the ideal expectation value is $\langle \bar{X}_0 \bar{X}_1 \rangle = 1$. The logical state preparation of $\ket{0}^{\otimes k}$ is treated as error free. Between each logical gate, physical qubits are subjected to independent depolarizing noise of rate $p$.  During syndrome extraction, bit-flips occur on the measured qubit with probability $p$. As the code is extended, the error in the expectation value of the observable decreases dramatically.

\begin{figure}
    \centering
    \includegraphics[width=\linewidth]{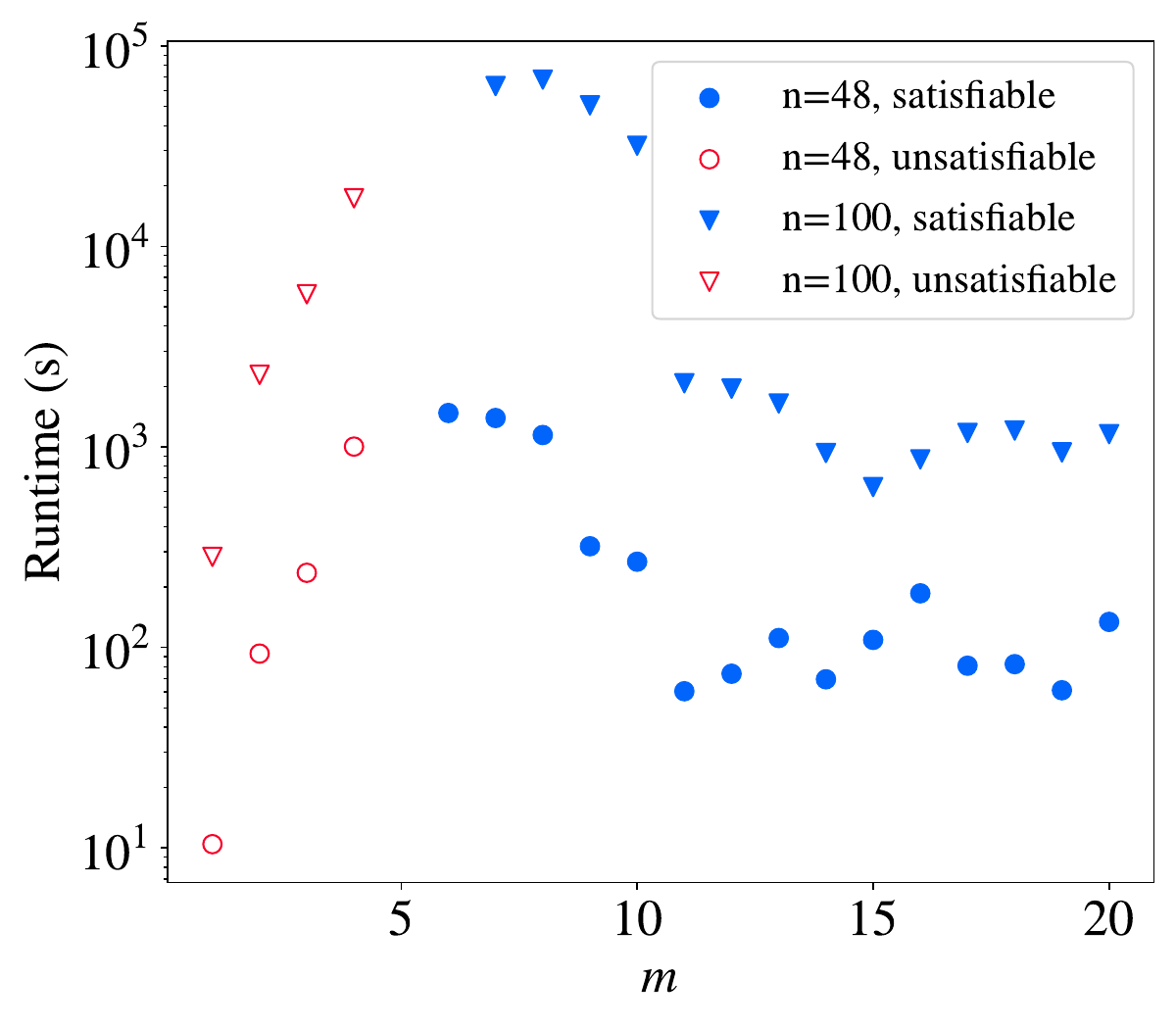}
    \caption{Time to extend the Fermi-Hubbard code~\eqref{eq:fh-g} from distance $d = 1$ to distance $d = 3$ vs. the number of qubits/stabilizers $m$ added. Timing is shown for both $n = 48$ and $n = 100$ physical qubits. As discussed in Sec.~\ref{sec:phase_transition} and demonstrated in Fig.~\ref{fig:phase_transition}, runtime peaks near the SAT--UNSAT phase transition, which here occurs around $m = 5$. All $n = 48$ code extensions run in at most twenty minutes of CPU time. Most $n = 100$ code extensions run in around 20 minutes, while the hardest cases take around a day.}
    \label{fig:fh-extension-time-n48-100}
\end{figure}

These results demonstrate the utility of physics-inspired code design and motivates the design of even larger codes. To demonstrate this, we next extend to $n = 48$ and $n = 100$ Fermi-Hubbard codes from distance one to distance three. The results are shown in Fig.~\ref{fig:fh-extension-time-n48-100}. As mentioned in the discussion of prior work in Sec.~\ref{sec:intro}, the problem size of $n = 100$ has become somewhat of a target in recent literature --- e.g., Ref.~\cite{olle_simultaneous_2024} presents numerical results for $n = 25$ qubits and a roadmap to $n = 100$ qubits with reinforcement learning, and Ref.~\cite{guerrero_game-theoretic_2026} uses a multi-agent system to optimize quantum codes and enable discovery of a $[\![100,50,4]\!]$ code in one hour. Here, our SAT framework can easily handle the more general problem of physics-informed code design for the $n = 100$ qubit target in just minutes to hours as shown in Fig.~\ref{fig:fh-extension-time-n48-100}. Indeed, most code design instances with $m \ge 10$ run in just minutes, while the hardest instances around $5 \le m \le 10$ take up to a day. This is in line with the phase transition in satisfiability/hardness demonstrated in Sec.~\ref{sec:phase_transition}, and we see that instances with $m \le 5$ turn out to be unsatisfiable. As a more direct comparison to Ref.~\cite{guerrero_game-theoretic_2026}, we also applied our code design formulation to the same $[\![100,50,4]\!]$ code \textit{discovery} problem, and found a satisfiable instance in under five minutes of CPU time compared to an hour in Ref.~\cite{guerrero_game-theoretic_2026}. This example is included under \textit{Code and Data Availability} in Sec.~\ref{sec:conclusion}.

\section{Hardware-Aware Code Design}
\label{sec:hardware-aware-code-design}

The formulation of code extension in terms of a satisfiability problem, outlined in Section~\ref{sec:sat}, is designed to find $m$ new stabilizer generators such that all errors in a given set can be detected. Given some maximal number of new qubits $m$ to be added when extending a code, it is not always possible to make every error in the set detectable. Instead, one would want to be able to detect as many errors as one can with the number of additional qubits (and thus extra stabilizers) allotted. This problem is an instance of maximum satisfiability, where the constraints that a given error is detectable are soft constraints, and the constraints that the group is abelian are treated as hard constraints. Given a hardware error model, a weight can be assigned to the clauses for each error informed by that model. A solution to this maximum satisfiability problem will then give a code that can detect as many of the errors as possible, prioritizing errors that are more likely on a given device.

On a given quantum device, certain Pauli errors will occur more frequently than others. It is common to consider an error model where the rate of Pauli $Z$ errors differs from that of Pauli $X$ and $Y$ errors. If $p_X$, $p_Y$, and $p_Z$ are the probabilities of $X$, $Y$, and $Z$ errors on any given qubit, respectively, then this model is parameterized by a bias parameter
\begin{equation}
    \label{eq:bias-parameter-eta}
    \eta = \frac{p_Z}{p_{XY}}
\end{equation}
where $p_{XY} = p_X = p_Y$.
If, for example, a $Z$-type error is much more likely than an $X$- or $Y$-type error, then the algorithm is likely to produce stabilizers that contain more $X$-type stabilizers.

\subsection{Hardware-Aware Fermi-Hubbard Code Extensions}
To demonstrate tailoring a code to biased noise in this setting, an extension to the Fermi-Hubbard code with stabilizer generators defined in Eq.~\eqref{eq:fh-g} is constructed, where the extended code either is or is not tailored to the bias described above. Two extra qubits and stabilizers are added to allow for this extension. For some fixed value of $p_X = p_Y$, we sweep over values of $\eta$, giving $p_Z = \eta p_X$. 

\begin{figure}
    \centering
    \includegraphics[width=\linewidth]{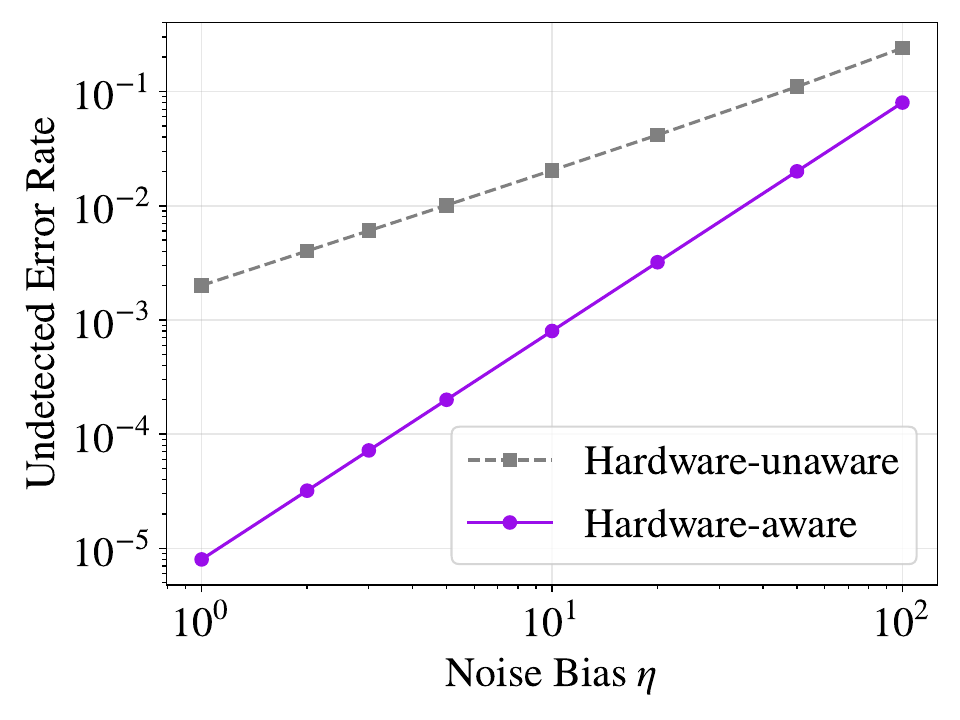}
    \caption{Sum of the rates of all undetectable weight-two errors versus the noise bias parameter $\eta$ for an extension to the $n=6$ Fermi-Hubbard code with $m=2$ added qubits. As shown, the hardware-aware code has a significantly lower undetected error rate than the hardware-unaware code for all noise biases. The overall rate of errors increases as the noise bias increases, so the two codes converge in the limit of large $\eta$.
    }
    \label{fig:bias-tailoring}
\end{figure}

Let $P = \bigotimes_{i=1}^n \sigma_i$ 
be an $n$-qubit error, where $\sigma_i \in \{I, X, Y, Z\}$. We compute the rate of the error as 
\begin{equation}
    r = \prod_{i=1}^n r_i,
\end{equation}
where $r_i = p_Z$ if $\sigma_i = Z$, $r_i = 1-p_Z-2p_X$ if $\sigma_i = I$, and $r_i = p_X$ otherwise. 
In Fig.~\ref{fig:bias-tailoring}, we compare the total rate of undetectable weight-two errors as a function of the noise bias $\eta$ for codes that are tailored and not tailored to the underlying noise.
For the hardware-aware code, the weights of the errors are the same as in the biased error model. For the hardware-unaware code, uniform weights are given to each error. As can be seen, the hardware-aware code has a lower undetected error rate, lower by as much as two orders of magnitude for small noise biases. This demonstrates the ability of our code design framework to adapt to given error models to improve error detection/correction under realistic hardware conditions.

\subsection{Hardware-Aware Surface Codes}
Next, we consider designing surface codes under biased noise. 
For a biased noise model, it is known that the XZZX surface code has a higher threshold than the rotated CSS surface code~\cite{bonilla_ataides_xzzx_2021}. The XZZX surface code is derived from the CSS surface code by applying Hadamard gates on alternating qubits, which results in stabilizers of the form $XZZX$ (or $XZ$ on the boundaries). In general, there are many codes that share the locality properties of the surface code while potentially being more suitable for certain noise models. Here we develop a technique to search for such codes using our SAT formulation.

\begin{figure}
    \centering
    \includegraphics[width=
    \linewidth]{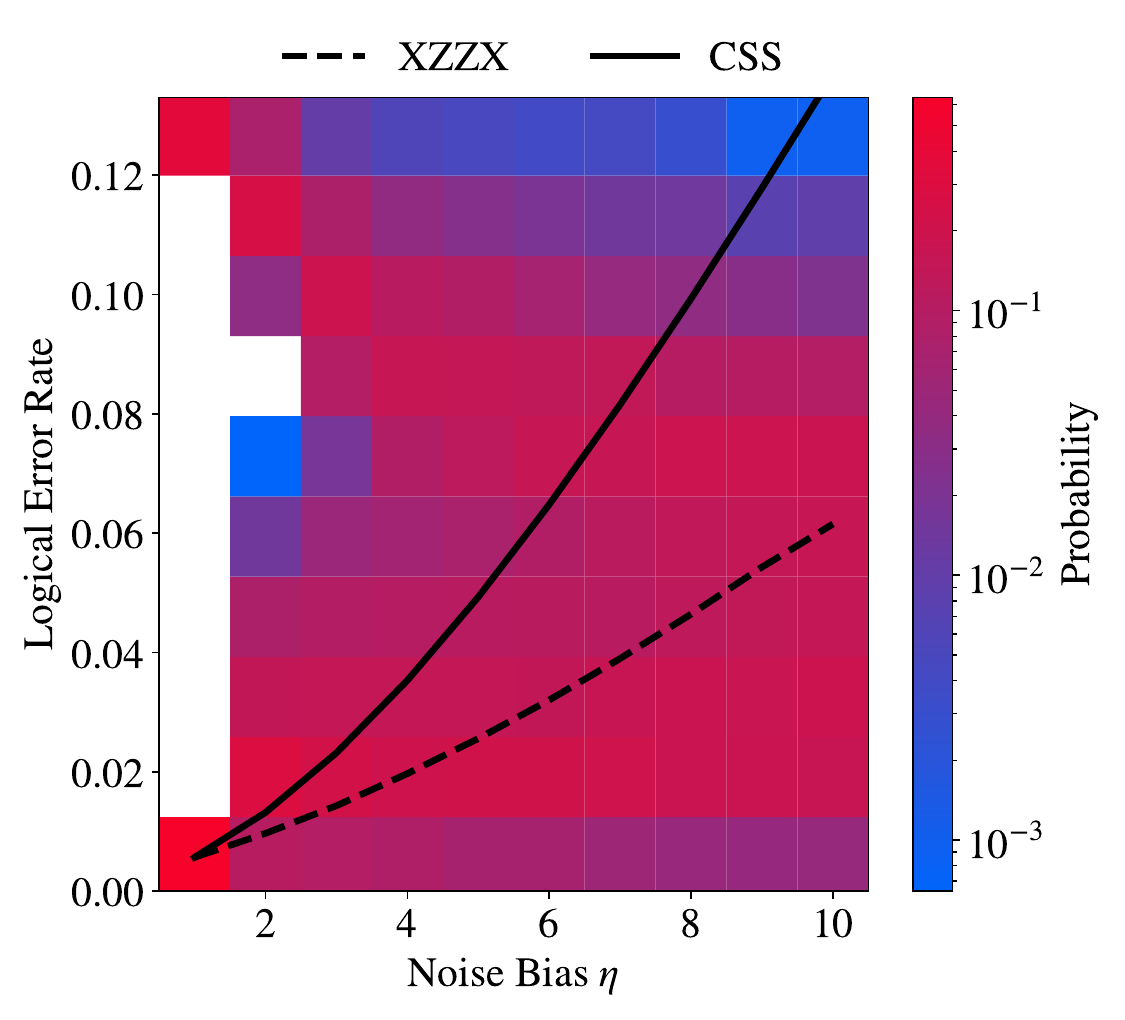}
    \caption{Distribution of logical error rate in designed surface codes vs. the bias parameter $\eta = p_X / p_{YZ}$. The CSS and XZZX codes are shown overlaying the designed codes. Here, $10^5$ designed codes are sampled uniformly at random from solutions to the SAT problem, and the colored square show frequencies (probabilities) of codes with the given logical error rate for the given noise bias $\eta$. (White squares represent zero.) Each column of the plot represents a probability distribution over the logical error rate for some value of $\eta$. As can be seen, many designed codes have a lower logical error rate than the XZZX surface code, which is a current state-of-the-art solution for biased noise.}
    \label{fig:ler-vs-eta-heatmap}
\end{figure}

To discover codes with the same locality as the surface code, each stabilizer is assigned some set of physical qubits where it can act non-trivially. In a rotated surface code, each physical qubit is assigned to a vertex of a graph, and each stabilizer corresponds to a face of the graph. When discovering surface-like codes, each unknown stabilizer generator is assigned to a face of the graph. A constraint is then imposed such that all qubits which do not correspond to vertices on that face of the graph must be assigned as identity in that stabilizer generator. If $F_i$ is the $i$\textsuperscript{th} face of the graph and $u_i = [u_i^X|u_i^Z]$ is the binary symplectic form of the $i$\textsuperscript{th} unknown stabilizer, then the constraint function is
\begin{equation}
    \bigwedge_{l \notin F_i} \neg(u_{i[l]}^X \vee u_{i[l]}^Z).
    \label{eq:surface-code-constr}
\end{equation}
All other qubits are constrained to be the identity (both the $X$ and $Z$ bits on that qubit are $0$). 

The constraint in Eq.~\eqref{eq:surface-code-constr} is imposed along with the usual constraints that the code must be abelian and that all errors up to the distance of the code must be detectable or in the stabilizer group. Using the SAT formulation, all codes that satisfy the constraints are seen as ``equal'' in that they can detect all desired errors and have an abelian stabilizer group. However, not all codes will have the same logical error rate under a given error model. 
Using an algorithm for approximately uniform sampling of SAT solutions, we can characterize the distribution of logical error rates among codes that share the same locality as the surface code but differ in their assignments of Pauli operators to the qubits. The CSS and XZZX surface codes are two codes from this distribution.

All codes discovered by sampling SAT solutions are subjected to a $Z$-basis memory experiment where the logical $\ket{0}$ state is prepared, the qubits are subjected to noise, and then the stabilizer generators and logical observable $\bar{Z}$ are measured. Decoding the syndromes gives the logical error rate, defined as the rate at which the decoder's predictions do not match the observed eigenvalue of the logical observable. Because the memory experiment in performed in the Pauli $Z$ basis, we choose to bias errors towards Pauli $X$. We choose an independent model in which Pauli $Y$ and $Z$ errors have some fixed, equal probability $p_{YZ}$ and the probability $p_X$ of an $X$ error varies. The bias is quantified by the parameter $\eta = p_X / p_{YZ}$. To find the lowest possible logical error rate that the code is capable of, a maximum likelihood decoder is used. The codes discovered by this method will generally not be CSS codes, and as such may not perform well with a minimum-weight perfect matching decoder.

Figure~\ref{fig:ler-vs-eta-heatmap} shows the logical error rate as a function of $\eta$ for the distance three CSS and XZZX surface codes, along with the distribution of the logical error rate at each value of $\eta$ for codes that were sampled from the SAT problem.  At low values of $\eta$, the distribution has a large peak near the CSS and XZZX codes, and a smaller peak at a high logical error rate. As $\eta$ increases, the peaks in the distributions move closer together. As can be seen, many codes perform better than the XZZX code, a current state-of-the-art surface code for biased noise. Interestingly, we find codes with logical error rates between the CSS and XZZX codes. These codes represent novel surface codes which, in a way, interpolate between the CSS and XZZX surface codes. 
%To see these advantage, we define the metric
% \begin{equation}
%     \label{eq:performance}
%     \bar{p}_C(\eta) = \frac{l_C(\eta) - l_\textrm{CSS}(\eta)}{l_\textrm{XZZX}(\eta) - l_\textrm{CSS}(\eta)},
% \end{equation}
% where $l_C(\eta)$ is the logical error rate of sampled code $C$ at noise bias $\eta$, and similarly for $l_\textrm{CSS}(\eta)$ and $l_\textrm{XZZX}(\eta)$. Around $\eta = 1$, where $l_\textrm{CSS}(\eta) = l_\textrm{XZZX}(\eta)$, we instead define
% \begin{equation}
%     \bar{p}_C(\eta) = \frac{l_C(\eta) - l_\textrm{XZZX}(\eta)}{l_\textrm{XZZX}(\eta)}
% \end{equation}
% Thus $\bar{p}_C(\eta) < 0$ if the code performs worse than the CSS code, and $0 \leq \bar{p}_C(\eta) \leq 1$ if the code performs better than the CSS code and worse than the XZZX code, and $\bar{p}_C(\eta) > 1$ if a code performs better than the XZZX code. Figure~\ref{fig:performance-distribution} shows the marginal distribution of $\bar{p}_C$ summed over $\eta$ for all sampled codes in Fig.~\ref{fig:ler-vs-eta-heatmap}. As can be seen, virtually all sampled codes outperform the CSS surface code, and a large number of sampled codes outperform the XZZX surface code. The codes with $0 \leq \bar{p}_C(\eta) \leq 1$ represent new codes which, in a way, interpolate between the CSS and XZZX surface codes. 
We note that while these codes have the same locality as the surface code, they may not guarantee efficient decoding. Nonetheless, codes with lower logical error rates than the XZZX code are particularly notable,  and advances in efficient maximum likelihood decoding may make the codes discovered here a practical choice for implementing error correction on devices with constrained connectivity between physical qubits. 

% \begin{figure}
%     \centering
%     \includegraphics[width=\linewidth]{performance_distribution.pdf}
%     \caption{%Distribution of the performance metric for the sampled surface codes. Each data point represents the performance (see Eqn.~\eqref{eq:performance}) of a single code averaged over the bias parameter $\eta$.
%     The performance metric $\bar{p}_C$ defined in Eq.~\eqref{eq:performance} for designed codes $C$ from Fig.~\ref{fig:ler-vs-eta-heatmap}, marginalized over all noise biases $\eta$. As can be seen, nearly all codes outperform the CSS surface code ($\bar{p}_C > 0$), while many codes outperform the XZZX surface code ($\bar{p}_C > 1$). Interestingly, we are able to design codes in the intermediate region $0 \le \bar{p}_C  \le 1$ which, in a way, interpolate between the CSS and XZZX surface code.}
%     \label{fig:performance-distribution}
% \end{figure}

\section{Conclusion}\label{sec:conclusion}

We have developed a general, flexible, and performant framework for quantum code design, analyzed its worst-case and average-case theoretical complexity, and demonstrated its utility by designing novel codes at scale with improved performance over state-of-the-art methods. In particular, we designed physics-inspired codes which incorporate symmetries of the Fermi-Hubbard model and demonstrated improved accuracy of observable expectation values in quantum error detection experiments. Additionally, we have designed hardware-aware surface codes which have a lower logical error rate than the XZZX surface code for biased noise. Beyond these examples, we present the framework to design additional quantum codes ``from scratch'' (code discovery) or from initial specifications (code extension) with the ability to incorporate realistic hardware noise. This ability will be crucial for experimentally demonstrating impactful applications of quantum computers. 

In this work, we restricted to abelian stabilizer codes in our examples, but our framework is extensible to alternative classes as well. Future work in designing entanglement-assisted codes~\cite{brun_correcting_2006}, subsystem codes~\cite{paolo2004,aly2006subsystemcodes}, and non-abelian codes will likely lead to additional impact. While we did not impose conditions on logical operators in this work, this can also be incorporated in future work. Current quantum error correction experiments are mostly memory experiments, where a state is prepared and then preserved through multiple rounds of error correction, but in  practical applications it is essential to perform logical operations with encoded qubits. Extending (or in general designing) codes with conditions on logical operations --- for example, transversal operations --- is the subject of future work. 
Finally, while we used our SAT formulation to design codes with improved logical error rates, this is an ``emergent'' property of the code that we did not incorporate into the design, but rather tested after. It is unlikely that one can easily incorporate the logical error rate into the SAT formulation or any formulation for code design since this is a property of the full code that must be measured by a separate (numerical) experiment. We leave as an open problem the possibility to directly incorporate logical error rate into the code design problem in future work.

We have proved general theoretical results that apply to all code design frameworks --- in particular, that the code design problem is NP-complete --- and we have demonstrated that our SAT formulation is capable of designing novel codes with advantages over current state of the art. While there has been tremendous progress in quantum hardware, a large-scale impactful application of an error corrected quantum computer remains a long sought-after milestone in the field. Designing quantum codes tailored to both the problem and the hardware can accelerate the timeline to this goal. Our work has already discovered novel advantageous codes, and our framework can continue to do so when applied to future problems. 

\textit{Code and data availability --} Our SAT formulation for code design is open source and freely available at~\cite{github}. This repository contains all examples and data to reproduce the results of this work.

\textit{Acknowledgments --} This work was supported in part through computational resources and services provided by the Institute for Cyber-Enabled Research at Michigan State University. B.D., W.M.W., M.L.L. and G.Q. acknowledge funding by the U.S. Department of Energy, Office of Science, Office of Advanced Scientific Computing Research, Accelerated Research in Quantum Computing under Award No. DE-SC0025509.

\bibliographystyle{apsrev4-1}
\bibliography{refs}

% \newpage
% \onecolumngrid
\appendix

\section{Subroutine to find the generators of the normalizer\label{app:normalizer}}

Finding the generators of the normalizer of a subgroup of Pauli group can be done in polynomial time. Take $\Sigma = \mathcal{N}(\mathcal{S})$, where
\begin{align}
    \mathcal{N}(\mathcal{S}) = \{ Q \in \mathcal{P}_n \mid [Q,S] = 0,\ \forall S \in \mathcal{S} \}.
\end{align}
Use the binary symplectic representation, so a Pauli string $S$ maps to the vector $s\in\mathbb{F}_2^{2n}$. Then commutation is
\begin{align}
    a \Lambda b^T = 0,
\end{align}
where 
\begin{align}
    \Lambda = \begin{pmatrix}
        0 & I_n \\
        I_n & 0
    \end{pmatrix}
\end{align}
Let 
\begin{align}
    S = \begin{pmatrix}
        s_1 \\
        s_2 \\
        \vdots \\
        s_m
    \end{pmatrix}
\end{align}
be the matrix representation the stabilizer set.
Then
\begin{align}
    S \Lambda q^T = 0
\end{align}
implies $Q$ is in the normalizer of $\mathcal{S}$. The space is
\begin{align}
    \mathcal{N}(\mathcal{S}) \cong \mathrm{ker}(S\Lambda)
\end{align}
The matrix $M \Lambda$ has dimension $m\times 2n$. Compute the rank $r$ of $M \Lambda$. Then $\mathrm{dim}\mathcal{N}(S) = 2n - r$. A basis can be found in polynomial time by Gaussian elimination, $\mathcal{O}(m n^2) \in \mathcal{O}(n^3)$, since the number of stabilizers is always less than the number of physical qubits, else the code space is trivial. Given the generators of the normalizer, products of them can be taken to generate the whole set of candidate Pauli strings.

\section{Hardness of finding code extensions which preserve the number of logical qubits} \label{sec:code-extensions-with-the-same-number-of-logical-qubits}

Theorem~\ref{thm:existence-of-distance-increasing-extensions} guarantees a distance-increasing extension; however, $k$ will increase under this procedure. As a minor corollary, we show that finding a concatenation that maintains the number of logical qubits reduces to finding a quantum code  with the added constraint that the code must encode exactly $k$ logical qubits.

\begin{corollary}
    Determining if there exists a concatenation of an $[\![n,k,d]\!]$ code $C_1$ with an outer code $C_2$ such that the resulting concatenated code $C_{1,2}$ is an $[\![n+m,k,d']\!]$ code with $d'>d$ and $k$ remaining the same reduces to solving an exact instance of $k$-\textsc{Code-Design} for the outer code.
\end{corollary}
\begin{proof}
    For this proof, we will define a specific procedure to create a concatenated code $C_{1,2}$ by enforcing certain properties on the outer code. Consider an $[\![n,k,d]\!]$ inner code $C_1$ and an outer code $C_2$ that is an $[\![n_2,k_2,d_2]\!]$ code. We require that $n_2 \operatorname{mod} k =0$ such that we can define a parameter $r \coloneq \frac{n_2}{k}$. 

    We define the concatenation procedure to begin by encoding $rn$ input registers via the $C_1$ code. This will result in $rk$ logical qubits of the $C_1$ code. Let the outer code, $C_2$ act on these logical qubits, such that $n_2 = rk$. Since the final number of logical qubits must remain the same, we require that this second code encode $k_2=k$ qubits.

    Finally, to ensure that the distance increases for the final code, we use the following relation for the distance $d'$ of a concatenated code constructed in this manner~\cite[Chp. 3]{Gaitan_2013} 
    \begin{equation}
        d' = \frac{d_1 d_2}{k_2} = \frac{d d_2}{k} \,. 
    \end{equation}
    Since we require $d' > d$, we require that $d_2 > k$. 

    Thus, $C_2$ must be a $[\![rk, k ,d_2]\!]$ code with $d_2 > k$. Note that for the simplest case where $d_2 = k+1$, the Singleton bound is satisfied as long as $r>1$. In general, the Singleton bound is satisfied for all $d_2 = k +q$ where $q \leq (r-2)k +1$. 

    Thus, the problem of extending an $[\![n,k,d]\!]$ code via concatenation such that its distance increases but the number of logical qubits remains constant can be solved by solving an exact instance of the $k$-\textsc{Code-Design} problem for an $[\![rk, k ,d_2]\!]$ code with $(r-1)k+1 \geq d_2 > k$.
\end{proof}

Such extensions always exist for $k=1$. For $k> 1$, the question is a bit trickier. For the $[\![4,2,2]\!]$ code, $r=2$ suffices with $C_2$ being the $[\![8,2,3]\!]$ Hermitian code~\cite{delfosse2020short}. If $C_1$ is the $[\![5,3,2]\!]$ code, $r=2$ would require a $[\![10,3,4]\!]$, which does not exist~\cite{Grassl:codetables,Grassl06,Magma,Brouwer98}; however, for $r=3$, there exists a $[\![15,3,5]\!]$ code that would suffice as a choice for $C_2$. Indeed, for a sufficiently large $r$, such an extension becomes increasingly likely. In fact, by the quantum Gilbert-Varshamov bound~\cite{Gottesman}, we can guarantee the existence of this extension by choosing $r$ satisfying
\begin{equation}
    \sum_{j=0}^k\binom{rk}{j}3^j\le2^{(r-1)k}.
\end{equation}

\section{Proof of code extension without mutation} \label{sec:bb-proof}

Here we prove Theorem~\ref{thm:bb-extension}. 
First we review the structure of bivariate bicycle (BB) codes as given in Ref.~\cite{bravyi_high-threshold_2024}. Let $S_\ell$ be the $\ell \times \ell$ shift matrix with entries $[S_{\ell}]_{ij} = \delta_{i, (j+1) \mod 2}$, where $\delta$ is the Kronecker delta. $S_\ell$ represents the cyclic permutation $(1,\ 2,\ \cdots,\ \ell)$. Let $\mathbb{I}_\ell$ be the $\ell \times \ell$ identity matrix. For some positive integers $\ell$ and $m$, define $x = S_\ell \otimes I_m$ and $y = \mathbb{I}_\ell \otimes S_m$. Let $A_1$, $A_2$, $A_3$, $B_1$, $B_2$, and $B_3$ each be a monomial in either $x$ or $y$. Then $A = A_1 + A_2 + A_3$ and $B = B_1 + B_2 + B_3$ are polynomials in $x$ and $y$. The BB code defined by the quadruple $(\ell, m, A, B)$ is the CSS code with $X$ and $Z$ check matrices $H^X=[A|B]$ and $H^Z=[B^T|A^T]$.

Let $x = S_\ell \otimes I_m$, $y = I_\ell \otimes S_m$, $x' = S_\ell \otimes I_{tm}$, and $y' = I_\ell \otimes S_{tm}$. Note $x,y \in \Ftwo^{\ell m \times \ell m}$ and $x',y' \in \Ftwo^{t \ell m \times t \ell m}$. For some integer exponent $p$, $x^p = S_\ell^p \otimes I_m$ and $y^p = I_\ell \otimes S_m^p$. Similarly, $(x')^p = S_\ell^p \otimes I_{tm}$ and $(y')^{tp} = I_\ell \otimes S_{tm}^{tp}$.

Our first goal is to construct a matrix $\txp \in \Ftwo^{\ell m \times t \ell m}$ from $x^p$ s.t. every row of $\txp$ also occurs as a row of $(x')^p$. For every element $(x^p)_{ij}$, set $\txp_{i,tj} = (x^p)_{ij}$. All other elements of $\txp$ are $0$. This will be called the ``padding'' of the matrix. By the definition of the tensor product,
\begin{equation}
    x^p = S^p_\ell \otimes I_m =
    \begin{bmatrix}
        [S^p_\ell]_{11} \mathbb{I}_m & \cdots & [S^p_\ell]_{1 \ell} \mathbb{I}_m \\
        \vdots & \ddots & \vdots \\
        [S^p_\ell]_{\ell 1} \mathbb{I}_m & \cdots & [S^p_\ell]_{\ell \ell} \mathbb{I}_m \\
    \end{bmatrix}
\end{equation}
and
\begin{equation}
    (x')^p = S^p_\ell \otimes I_{tm} =
    \begin{bmatrix}
        [S^p_\ell]_{11} \mathbb{I}_{tm} & \cdots & [S^p_\ell]_{1 \ell} \mathbb{I}_{tm} \\
        \vdots & \ddots & \vdots \\
        [S^p_\ell]_{\ell 1} \mathbb{I}_{tm} & \cdots & [S^p_\ell]_{\ell \ell} \mathbb{I}_{tm} \\
    \end{bmatrix}
    .
\end{equation}
Since all rows and columns of $S^p_\ell$ have Hamming weight $1$, all rows and columns of $x^p$ and $(x')^p$ also have Hamming weight one. By construction, all rows of $\txp$ have Hamming weight one and all columns have Hamming weight either one or zero. The rows of all of these matrices are unique. For a row of $\txp$ to equal a row of $(x')^p$, it suffices to check that they have their $1$ in the same column. If $[S^p_\ell]_{ij} = 1$, then for any $k = 1, \ldots m$,
\begin{equation}
    \label{eq:xp-one-index}
    [x^p]_{(i-1)m+k, (j-1)m+k} = 1 \\
\end{equation}
and
\begin{equation}
    \label{eq:txp-one-index}
    [\txp]_{(i-1)m+k, t((j-1)m+k)} = 1.
\end{equation}
For any $k = 1, \ldots, tm$,
\begin{equation}
    \label{eq:xpp-one-index}
    [(x')^p]_{(i-1)tm+k, (j-1)tm+k} = 1.
\end{equation}
For some $k$ between $1$ and $m$, the row of $\txp$ with a $1$ in column $t((j-1)m+k)$ is equal to a row of $(x')^p$ with column index $(j-1)tm + k'$ for some $k'$ with $1 \leq k' \leq tm$. Equating the two column indices, $k'=tk$. Thus the row of $\txp$ with index $(i-1)m+k$ is equal to the row of $(x')^p$ with index $(i-1)tm + k' = t((i-1)m + k)$. In words, for any row $i$ of $\txp$, row $ti$ of $(x')^p$ is equal to it.

Next construct $\typ$ from $y^p$ in the same manner as for $\txp$. It must be shown that every row of $\typ$ is also a row of $(y')^{tp}$. By definition,
\begin{equation}
    y^p = I_\ell \otimes S_m^p =
    \begin{bmatrix}
        S_m^p & & \\
        & \ddots & \\
        & & S_m^p \\
    \end{bmatrix}
\end{equation}
and
\begin{equation}
    (y')^{tp} = I_\ell \otimes S_{tm}^{tp} =
    \begin{bmatrix}
        S_{tm}^{tp} & & \\
        & \ddots & \\
        & & S_{tm}^{tp} \\
    \end{bmatrix}
    .
\end{equation}
Suppose $[y^p]_{ij} = 1$. Then $\typ_{i,tj} = 1$. By the definition of the shift matrices, $i = j - p \mod \ell$. If $[(y')^{tp}]_{i',j'} = 1$, then similarly $i' = j' - p \mod \ell$. Setting $j' = tj$, $i' = tj - p \mod \ell$ and $i' = ti$. Thus the $i$\textsuperscript{th} rows of $\txp$ and $\typ$ equal the $ti$\textsuperscript{th} rows of $(x')^p$ and $(y')^{tp}$.

Let $A(x,y) = x^{p_1} y^{q_1} + x^{p_2} y^{q_2} + x^{p_3} y^{q_3}$. The terms in $A$ must be monomials, so $q_i = 0$ if $p_i \neq 0$ and vice versa. Let $B(x,y) = x^{r_1} y^{s_1} + x^{r_2} y^{s_2} + x^{r_3} y^{s_3}$. For every monomial $x^{p_i} y^{q_i}$ in $A$, the $i$\textsuperscript{th} row of padding of the monomial corresponds to the $ti$\textsuperscript{th} row of the monomial $(x')^{p_i} (y')^{t q_i}$ in $A'$. The same condition applies to $B$ and $B'$. Since this correspondence does not depend on the exponents $p_i$ or $q_i$ ($r_i$ or $s_i$), the corresponding rows between the two matrices can be summed over all three monomials and the $i$\textsuperscript{th} row of $A$ ($B$) corresponds to the $ti$\textsuperscript{th} row of $A'$ ($B'$). The $X$-type check matrices of $C$ and $C'$ are $H^X = [A|B]$ and $(H^X)'=[A'|B']$, respectively. The $i$\textsuperscript{th} row of $H^X$ corresponds to the $ti$\textsuperscript{th} row of $(H^X)'$.

Now consider the $Z$-type check matrix $H^Z = [B^T|A^T]$ for the original BB code along with its counterpart $(H^Z)'=[(B')^T|(A')^T]$ for the larger BB code. We must show that for every row of $B^T$ ($A^T$) its padding is also a row of $(B')^T$ ($(A')^T$).

Let $\widetilde{((x^p)^T)}$ be the padding of $(x^p)^T$. Every row and column of $(x^p)^T$ has weight one, as do the rows and columns of $((x')^p)^T$. Since the rows and columns of $(x^p)^T$ have weight one, so do the rows and columns of $\widetilde{((x^p)^T)}$. Analogous to Eqs.~\eqref{eq:xp-one-index} through \eqref{eq:xpp-one-index}, for $k=1,\ldots,m$
\begin{equation}
    [(x^p)^T]_{(j-1)m+k, (i-1)m+k, } = 1 \\
\end{equation}
and
\begin{equation}
    [(\widetilde{(x^p)^T)}]_{(j-1)m+k, t((i-1)m+k)} = 1.
\end{equation}
For any $k = 1, \ldots, tm$,
\begin{equation}
    [((x')^p)^T]_{(j-1)tm+k, (i-1)tm+k} = 1.
\end{equation}
Since the rows and columns of $\widetilde{((x^p)^T)}$ have weight one as do the rows and columns of $((x')^p)^T$, a row of $\widetilde{((x^p)^T)}$ is equal to a row of $((x')^p)^T$ if they have a one in the same location. For any $k = 1, \ldots, m$ there must exist some $k' = 1, \ldots, tm$ s.t.
\begin{equation*}
    t((i-1)m+k) = (i-1)tm+k',
\end{equation*}
which is satisfied by $k'=tk$. Thus for every row of $(x^p)^T$, its padding is also a row of $((x')^p)^T$.

Let $\widetilde{((y^p)^T)}$ be the padding of $(y^p)^T$. Recall that $(y^p)^T = I_\ell \otimes (S^p_m)^T$ and $((y')^{tp})^T = I_\ell \otimes (S^{tp}_{tm})^T$. Written out explicitly,
\begin{equation}
    (y^p)^T = I_\ell \otimes S_m^p =
    \begin{bmatrix}
        (S_m^p)^T & & \\
        & \ddots & \\
        & & (S_m^p)^T \\
    \end{bmatrix}
\end{equation}
and
\begin{equation}
    (y')^{tp} = I_\ell \otimes S_{tm}^{tp} =
    \begin{bmatrix}
        (S_{tm}^{tp})^T & & \\
        & \ddots & \\
        & & (S_{tm}^{tp})^T \\
    \end{bmatrix}
    .
\end{equation}
Suppose $[(y^p)^T]_{ij} = 1$. Then $[\widetilde{((y^p)^T)}]_{i,tj} = 1$. By the definition of the shift matrices and the transpose that was performed, $j = i - p \mod \ell$. If for some $(i',j')$ $[(y')^{tp}]_{i',j'} = 1$, then similarly $j' = j' - p \mod \ell$. Setting $j' = tj$, $i' = tj - p \mod \ell$ and $i' = ti$. Thus the $i$\textsuperscript{th} rows of $\widetilde{((x^p)^T)}$ and $\widetilde{((y^p)^T)}$ equal the $ti$\textsuperscript{th} rows of $((x')^p)^T$ and $((y')^{tp})^T$. Since the transpose in $A^T$ and $B^T$ distributes over the monomial terms, the rest of the proof proceeds as for $H^X$ above, and for every row of $H^Z$ we find that its padding is also a row of $(H^Z)'$.

This construction requires that columns and rows of the original check matrix become spaced out by an interval $t$ when going from $H^X$ to $(H^X)'$. To restore the standard form of the check matrix for an extended code, permute the rows and columns of $(H^X)'$ s.t. $H^X$ is in its upper-left block. Permuting the rows does nothing to the code, and permuting the columns (\textit{i.e.} physical qubits) does not affect the distance of the code.

\section{More Details on the SAT--UNSAT Phase Transition\label{app:phase_transition}}

\begin{figure*}
    \centering
    \includegraphics[width=0.9\linewidth]{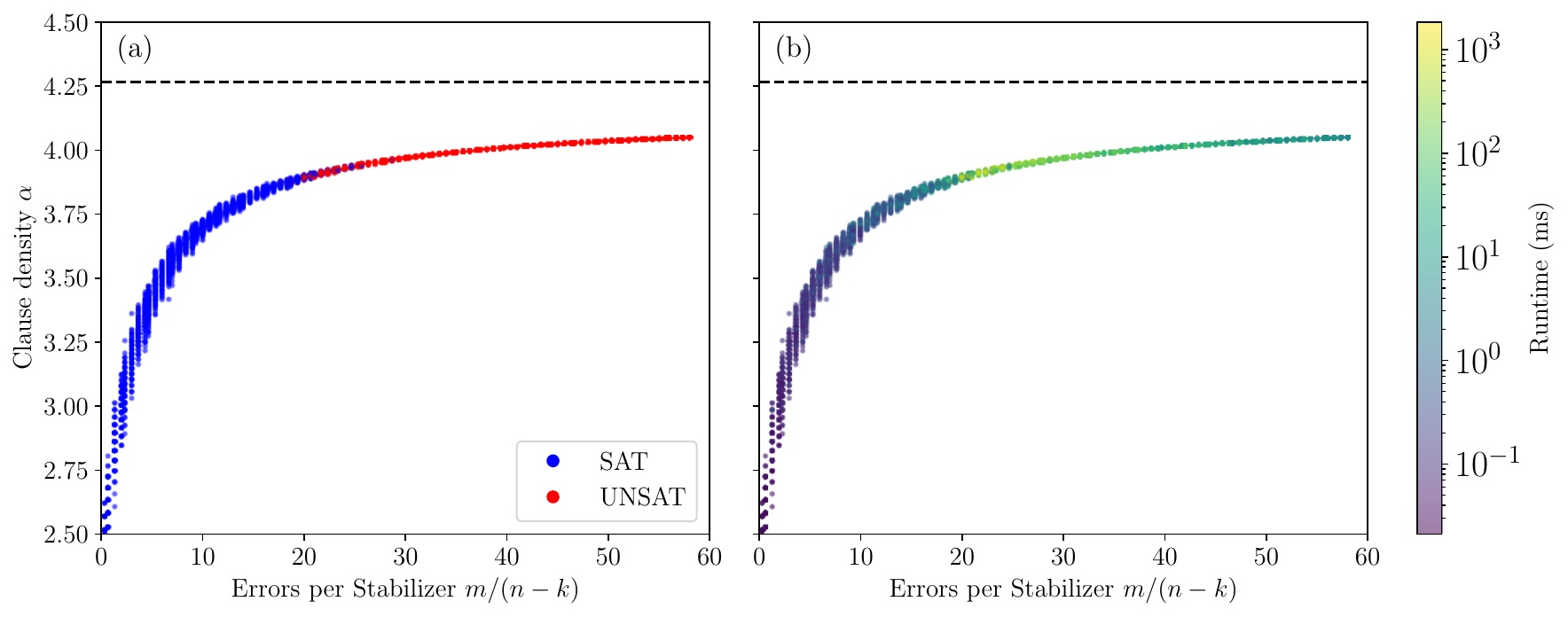}
    \caption{
    Relationship between the natural code-design density $m/(n-k)$ and the clause density $\alpha$ of the corresponding SAT instance for $n=4$, $k=1$, and $w=3$.
    The panel (a) indicates whether each  random instance is satisfiable (SAT) or unsatisfiable (UNSAT), while the panel (b) shows the corresponding SAT-solver runtime.
    The horizontal dashed line at $\alpha \simeq 4.267$ denotes the critical clause density of random 3-SAT given as a reference, since the code-design instances produce structured mixed 2--3-SAT formulas.
    }
    \label{fig:clause_density_vs_m}
\end{figure*}

\begin{figure*}
    \centering
    \includegraphics[width=0.75\linewidth]{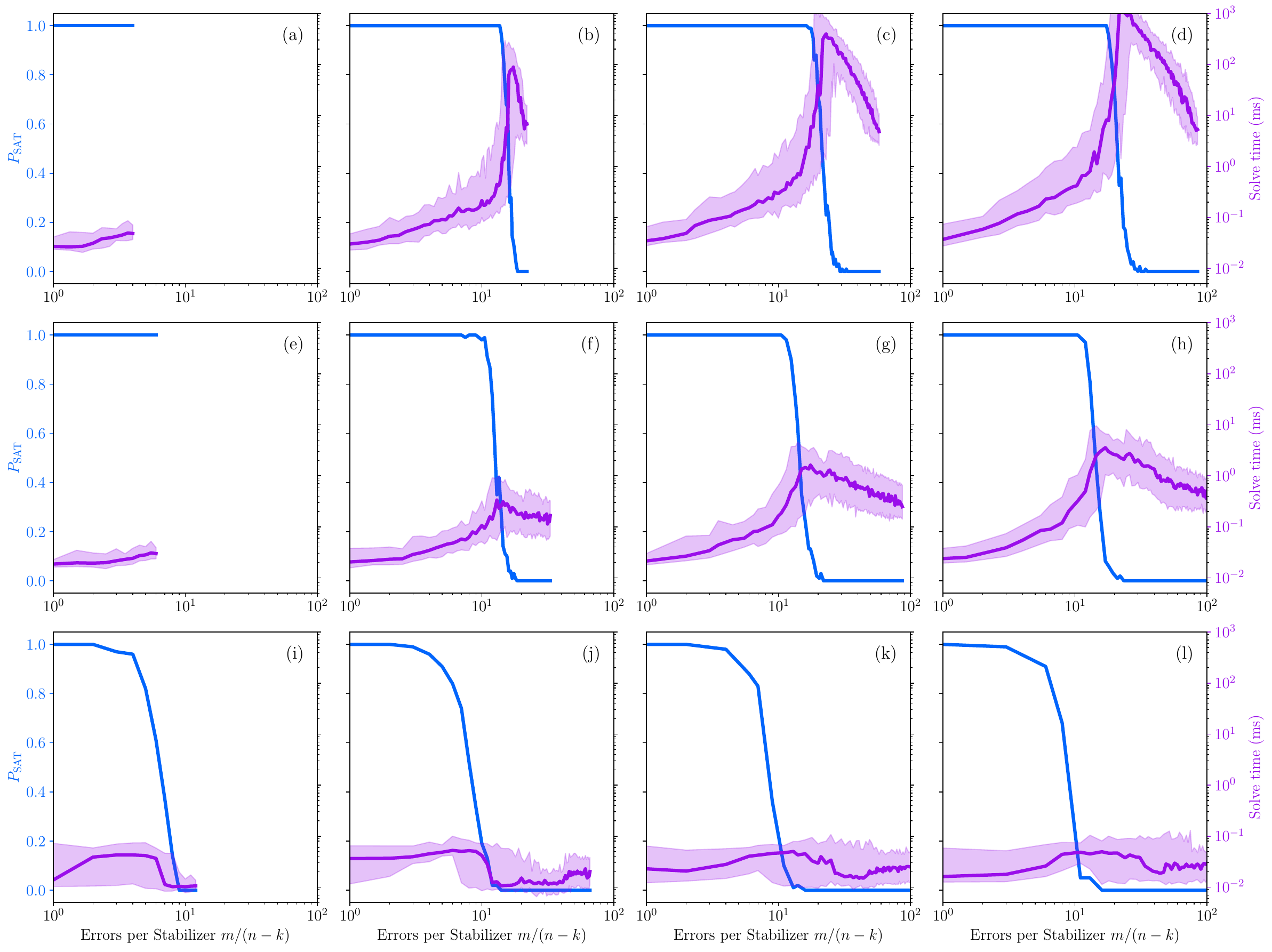}
    \caption{
    Finite-size SAT--UNSAT transitions for random code-design instances with $n=4$.
    Rows correspond to increasing numbers of logical qubits $k$, and columns correspond to increasing maximum error weights $w$.
    The blue curves show the empirical satisfiability probability $P_{\mathrm{SAT}}$ as a function of the code-design density $m/(n-k)$.
    The purple curves show the median SAT-solver runtime, with the shaded region indicating the 90\% confidence interval.
    The characteristic increase in runtime near the SAT--UNSAT crossover illustrates the easy--hard--easy behavior associated with NP-hard problems.
    }
    \label{fig:phase_transition_array}
\end{figure*}

To characterize the finite-size SAT--UNSAT transition systematically, we generate random $k$-\textsc{Code Design} instances for fixed $(n,k,w)$, where $w$ is the maximum Pauli weight included in the error pool, and vary the number of errors samples, $m$, from this pool.
For each parameter point, we generate 100 independent instances and estimate the satisfiability probability $P_{\mathrm{SAT}}$.
Figure~\ref{fig:clause_density_vs_m} illustrates the relation between the resulting SAT instances, their clause density, and the natural code-design density $m/(n-k)$ for $n=4$, $k=1$, and $w=3$.
The dashed line at $\alpha \simeq 4.267$ denotes the critical clause density of random 3-SAT and is included only as a reference, since the formulas generated by $k$-\textsc{Code Design} are structured mixed 2--3-SAT instances.

Figure~\ref{fig:phase_transition_array} shows the same transition over a broader set of parameters for $n=4$, with rows corresponding to $k=1,2,3$ and columns to $w=1,2,3,4$. 
Figure~\ref{fig:phase_transition_array}(d) is reported in the main text as Fig.~\ref{fig:phase_transition} to demonstrate the spike in runtime complexity at the $P_\mathrm{SAT}$ phase transition.
The blue curves report the satisfiability probability $P_{\mathrm{SAT}}$, while the purple curves show the median SAT-solver runtime and the shaded region illustrates the 90\% confidence interval.
The location of the transition varies appreciably with $(n,k,w)$, demonstrating that $m/(n-k)$ alone does not define a universal critical density.

\begin{figure}
    \centering
    \includegraphics[width=\linewidth]{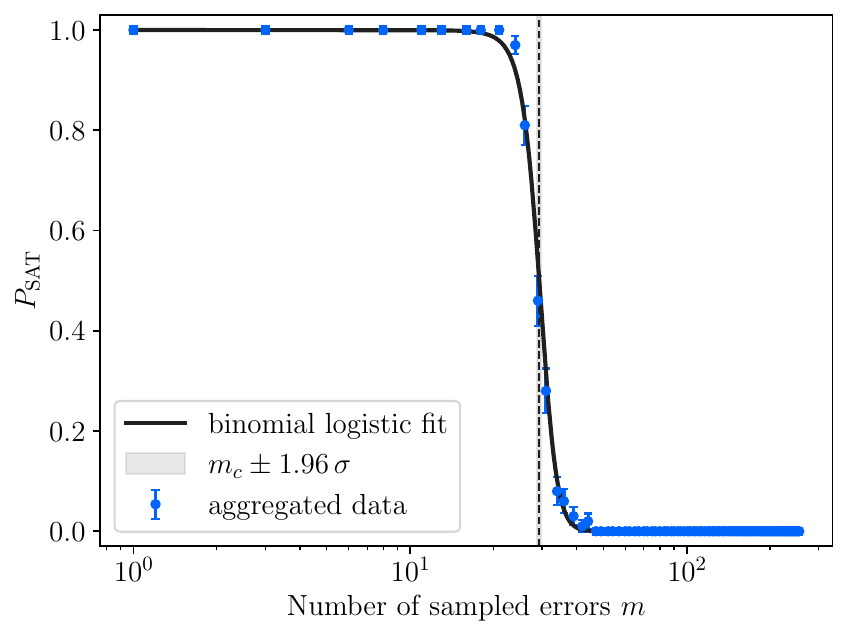}
    \caption{
    Example of random code-design instances fitted to extract the critical $m_c$ of the finite-size SAT--UNSAT transition.
    The empirical satisfiability probability $P_{\mathrm{SAT}}(m)$ is fit to a logistic function, with $m_c$ defined by the $P_{\mathrm{SAT}}=1/2$ crossing.
    The fitted width parameter characterizes the sharpness of the finite-size transition.
    }
    \label{fig:phase_transition_psat_fit}
\end{figure}

It is clear from Fig.~\ref{fig:phase_transition_array}, that while illustrative, the errors per stabilizer parameter [$m/(n-k)$] does not describe the order parameter since the critical value increases with increasing $n-k$.
Therefore, we extract a parameter-dependent critical point $m_c(n,k,w)$ by fitting each transition to
\begin{align}
    P_{\mathrm{SAT}}(m) = \frac{1}{ 1+\exp\left(\frac{m-m_c}{a}\right) },
\end{align}
where $m_c$ is the 50\% satisfiability crossing and $a$ determines the transition width, as illustrated in Fig.~\ref{fig:phase_transition_psat_fit}.
Of the 131 sampled $(n,k,w)$ curves, 46 include the SAT--UNSAT transition within the sampled range and are retained for the scaling analysis.
Defining the number of stabilizers $r=n-k$ and the number of Pauli errors of weight at most $w$ as
\begin{align}
    M_{\leq w}(n) = \sum_{j=1}^{w} 3^j {n \choose j},
\end{align}
a model comparison with the Bayesian information criterion (BIC) favors the empirical scaling form
\begin{align}
    m_c = c_0 + A n^a r^b\left[\log M_{\leq w}(n)\right]^ck^d w^e.
\end{align}
After model selection and refitting to all identifiable critical points, we obtain the fit parameters: $c_0=3.29$, $a=1.01$, $b=1.78$, $c=1.85$, $d=-0.34$, and $e=-0.44$. Note that the scaling law indicates a strong dependence on both $r$ and the number of Pauli errors.

\begin{figure}
    \centering
    \includegraphics[width=\linewidth]{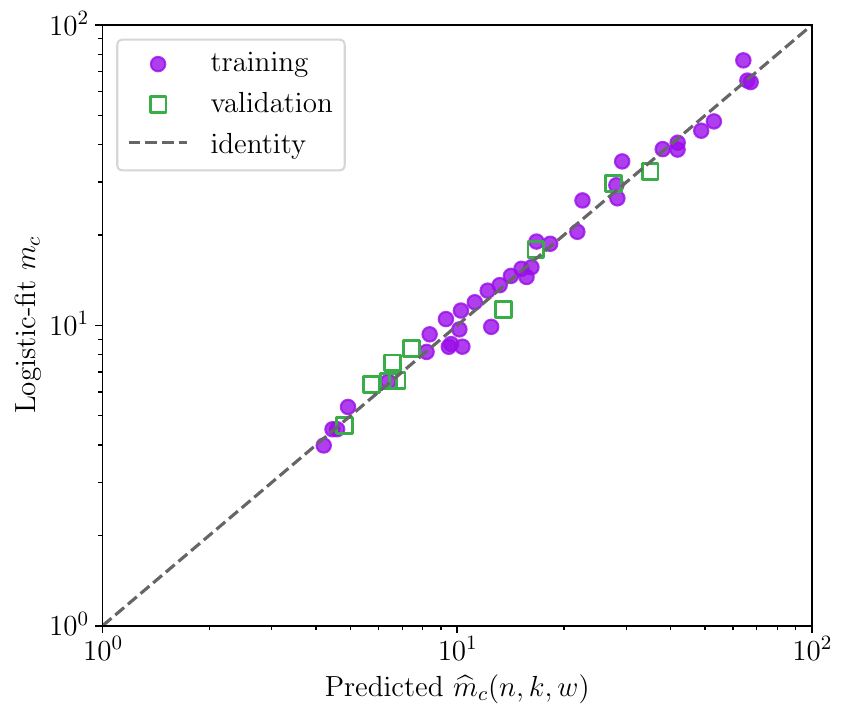}
    \caption{
    Comparison of the observed critical $m_c$, obtained from individual logistic fits, with the values predicted by the empirical finite-size scaling model.
    Agreement with the dashed $y=x$ line indicates accurate prediction of the location of the SAT--UNSAT crossover across different $(n,k,w)$ for training and validation sets.
    }
    \label{fig:phase_transition_pred_vs_obs}
\end{figure}

\begin{figure}[t]
    \centering
    \includegraphics[width=\linewidth]{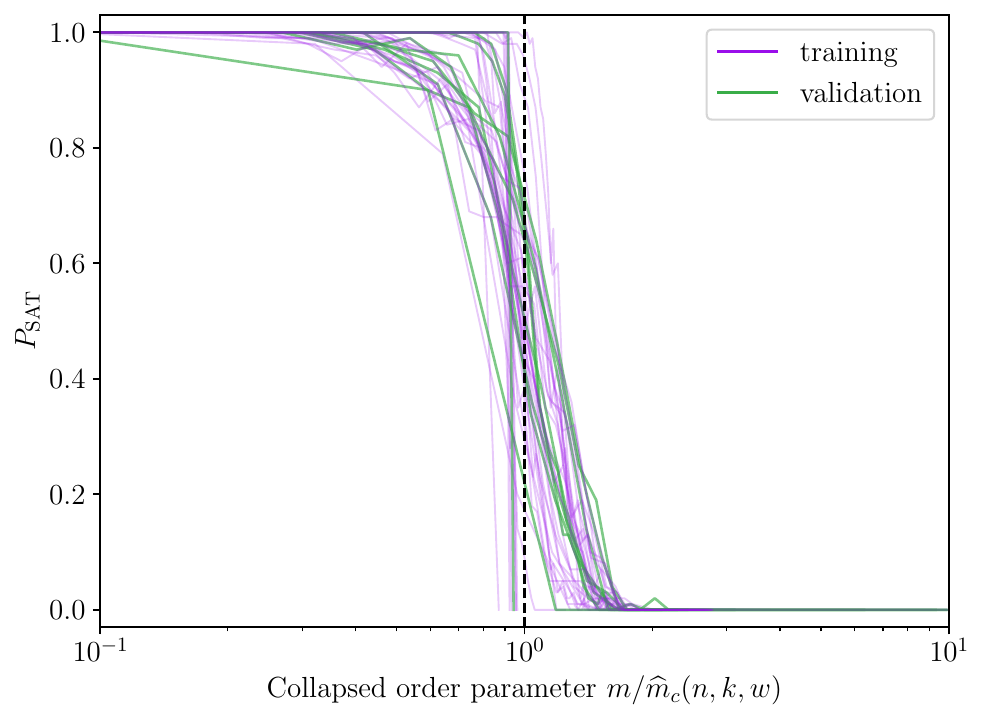}
    \caption{
    Collapse of the satisfiability transitions under rescaling by the predicted critical point $\widehat{m}_c(n,k,w)$.
    The collapse demonstrates that the empirical scaling model captures much of the parameter dependence of the finite-size transition.
    }
    \label{fig:phase_transition_data_collapse}
\end{figure}

The fit is performed in conjunction with cross-validation procedures. In particular, the critical $m_c$ dataset is split into training and validation sets. A fit using only the training points gives a held-out RMSE of $0.096$ in $\log m_c$, corresponding to an RMS relative error of approximately $9.7\%$, while five-fold cross-validation gives $0.105\pm0.040$ in $\log m_c$.
Figure~\ref{fig:phase_transition_pred_vs_obs} compares the predicted and fitted critical points, and Fig.~\ref{fig:phase_transition_data_collapse} shows the resulting collapse of the transition under the rescaled order parameter $m/\widehat{m}_c(n,k,w)$.
We emphasize that this relation is an empirical finite-size scaling ansatz over the sampled range rather than an asymptotic scaling law.

\newpage

\section{Additional Numerical Results and Discussion for Physics-Informed Code Design}

\begin{figure}
    \centering
    \includegraphics[width=\linewidth]{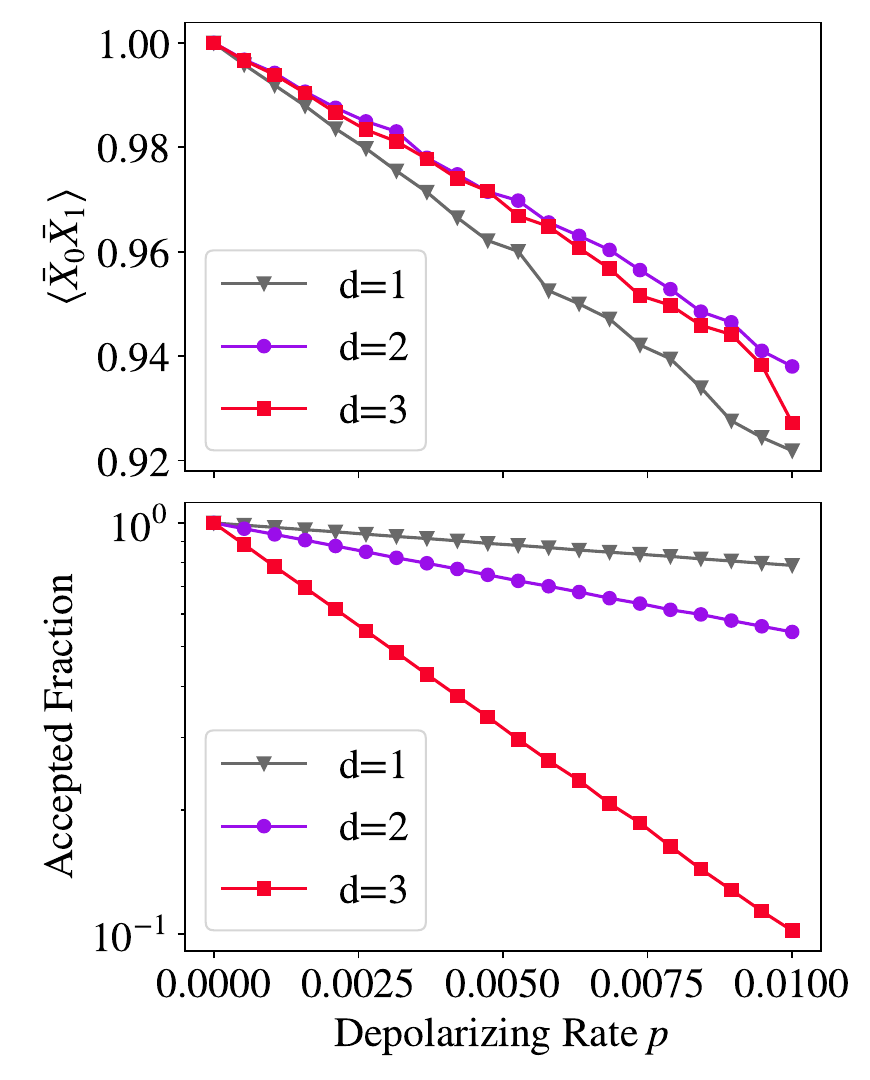}
    \caption{(top) Expectation value of logical observable $\bar{X}_0 \bar{X}_1$ in a logical Bell state vs. the depolarizing noise rate $p$ acting on the qubits for Fermi-Hubbard codes with distances 1, 2, and 3. The qubits are subject to circuit-level noise during the execution of each logical gate. (bottom) The fraction of shots that are accepted at each code distance vs. the depolarizing noise rate $p$. Points with no accepted shots are not shown.}
    \label{fig:fh-error-detection-realistic}
\end{figure}

With the simple noise model detailed in Section \ref{sec:physics-informed-code-design}, the result for the distance-3 code is almost perfect (see Fig.~\ref{fig:fh-error-detection}).
Fig.~\ref{fig:fh-error-detection-realistic} shows a similar simulation under a circuit-level noise model, where physical qubits experience noise after each physical gate in the gadgets implementing the logical circuits. While the improvement in expectation values in this case is, as expected, less emphatic than in Fig.~\ref{fig:fh-error-detection}, we still see that the extended codes perform better even when state preparation is noisy.

Because the SAT formulation only guarantees that errors are detectable, with few other constraints on the code, there is no guarantee that the logical operations of the resulting codes will have fault-tolerant implementations. This is the source of the failure of the code extension seen in Fig.~\ref{fig:fh-error-detection-realistic}. Extending a quantum code in such a way that
all fault-tolerant operations remain fault-tolerant
in the extended code will be the subject of future work. Indeed, there must be some sequences of extended codes at successively greater distances where operations remain transversal.

On fault-tolerant quantum hardware, QEC codes (\textit{e.g.} the surface code or QLDPC codes) will be implemented in order to perform computations with logical qubits. Even with very performant decoders, logical errors can still occur during the course of a computation. At the cost of a few extra logical qubits, our construction could be used to detect these logical errors in addition to detecting violations in the symmetries of the physical system (or other problem) being studied. Under these conditions, logical errors might be sufficiently rare that there is an advantage to employing the codes constructed here in order to enhance existing symmetry verification procedures with the ability to detect logical errors.

\begin{figure}[H]
    \centering
    \includegraphics[width=\linewidth]{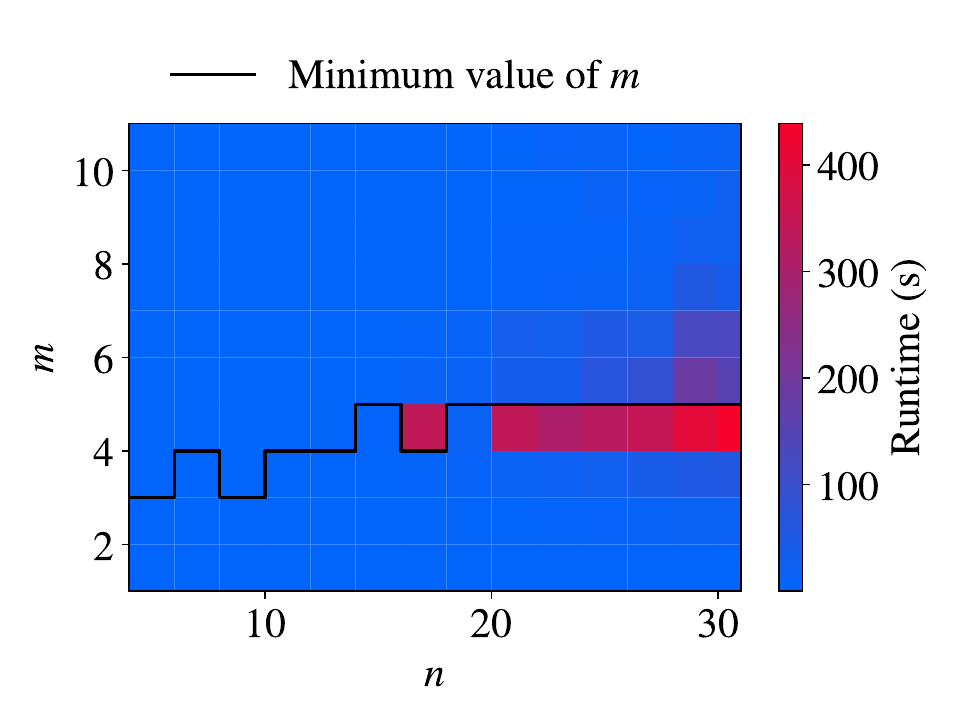}
    \caption{Time to use SAT to increase the FH code from distance 1 to distance 3 as a function of the number of qubits and stabilizers added to the code ($m$) and the number of qubits in the original system ($n$). The runtime indicates the time to either prove the problem is unsatisfiable for the given $m$ or find a distance-three extension. The black line shows the minimal value $m$ above which the problem is satisfiable.}
    \label{fig:fh-extension-time-with-m}
\end{figure}

To evaluate the performance of code extension, we consider the problem of starting with the original FH code with generators given in Eq.~\eqref{eq:fh-g} and extending the code to have distance three. The original code has stabilizer generators that consist of only Pauli $Z$ operators, and so only bit flip errors can be detected. A distance-three extension is capable of detecting any Pauli operators up to weight three. Fig.~\ref{fig:fh-extension-time-with-m} shows the time needed to solve this extension problem (or prove it is unsatisfiable) when adding $m$ new stabilizers to an $n$-qubit FH code. The black line shows the minimum value of $m$ that is necessary to increase the code to distance three. Many instances are seen to be easy, completing in around a minute. The hard instances, which take several minutes, lie along the satisfiable/unsatisfiable boundary for $n \geq 15$. This spike in the solution time is reminiscent of the phase transition seen in Sec.~\ref{sec:phase_transition}: when $m$ is small, the problem is easily proven unsatisfiable and when $m$ is large there are many degrees of freedom with which to create a code and the problem is easily satisfiable. The problems along the satisfiable/unsatisfiable boundary are the ones where the runtime required by the SAT solver may significantly increase. Similar results for larger systems can be seen in Fig.~\ref{fig:fh-extension-time-n48-100} of the main text.

\end{document}